\documentclass{article}

\usepackage{arxiv}
\usepackage{graphicx}
\usepackage[utf8]{inputenc} % allow utf-8 input
\usepackage[T1]{fontenc}    % use 8-bit T1 fonts
\usepackage{hyperref}       % hyperlinks
\usepackage{url}            % simple URL typesetting
\usepackage{booktabs}       % professional-quality tables
\usepackage{amsfonts}       % blackboard math symbols
\usepackage{nicefrac}       % compact symbols for 1/2, etc.
\usepackage{microtype}      % microtypography
\usepackage{lipsum}
\usepackage{color}
\usepackage{amsmath}

\newcommand{\graft}{\emph{GRAFT }}
\newcommand{\hgraft}{\emph{HGRAFT }}

\title{It Runs in the Family: Searching for Synonyms Using Digitized Family Trees}

\author{
  Aviad Elyashar, Rami Puzis, and Michael Fire \\
  Department of Software and Information Systems Engineering,\\
  Ben-Gurion University of the Negev, Beer-Sheva, Israel \\
  \textit{aviade@post.bgu.ac.il, \{puzis, mickyfi\}@bgu.ac.il} \\
}

\begin{document}
\maketitle

\begin{abstract}
Searching for a person’s name is a common online activity.
However, Web search engines provide few accurate results to queries containing names.
In contrast to a general word which has only one correct spelling, there are several legitimate spellings of a given name. 
Today, most techniques used to suggest synonyms in online search are based on pattern matching and phonetic encoding, however they often perform poorly.
As a result, there is a need for an effective tool for improved synonym suggestion. 
In this paper, we propose a revolutionary approach for tackling the problem of synonym suggestion.
Our novel algorithm, \emph{GRAFT}, utilizes historical data collected from genealogy websites, along with network algorithms.
\graft is a general algorithm that suggests synonyms using a graph based on names derived from digitized ancestral family trees.
Synonyms are extracted from this graph, which is constructed using generic ordering functions that outperform other algorithms that suggest synonyms based on a single dimension, a factor that limits their performance.    
We evaluated \emph{GRAFT}'s performance on three ground truth datasets of forenames and surnames, including a large-scale online genealogy dataset with over 16 million profiles and more than 700,000 unique forenames and 500,000 surnames.
We compared \emph{GRAFT}'s performance at suggesting synonyms to 10 other algorithms, including phonetic encoding, string similarity algorithms, and machine and deep learning algorithms.  
The results show \emph{GRAFT}'s superiority with respect to both forenames and surnames and demonstrate its use as a tool to improve synonym suggestion. 

\end{abstract}

\newcommand{\redcomment}[1]{{\color{red}{#1}}}
\newcommand{\remove}[1]{}

\keywords{Synonym Suggestion \and Digitized Family Trees \and Networks \and Network Science \and Personal Names \and Name-Based Graphs}

\section{Introduction}
\label{sec:introduction}

Searching for a person's name is a frequent activity in information systems~\cite{yang2006web}.
Retrieving a news article by using the author's name, examining patient records~\cite{pfeifer1996retrieval}, and finding usernames via received emails~\cite{minkov2006contextual} are all daily activities performed using individuals' names. 
Moreover, dependency on names for Web searches is growing. 
In 2004, 30\% of search engine queries included personal names~\cite{guha2004disambiguating}.
One decade later, one billion names were entered into the Google search engine
each day~\cite{GoogleYourself}.

While online searching for people's names has increased, the results from Web search engines have not kept pace. 
Well-known search engines like Google, Yahoo, and Bing provide few accurate results in response to queries containing names. 
This acute problem has created a new market need\footnote{\url{https://organicweb.com.au/social-media/people-search-pipl-wink-peekyou/}} %~\cite{organicweb} 
that has been filled by companies, e.g., Pipl,\footnote{https://pipl.com/}%~\cite{pipl} 
%and ZoomInfo\footnote{https://www.zoominfo.com/}%~\cite{zoominfo}%
which specialize in searching for information about specific people.
However, in many cases, users do not know the exact name or the correct form of the name that they are searching for. 
Therefore, searching online for people by their names remains a challenging problem.

\begin{figure}[h!]
\centering
\includegraphics[scale=0.7]{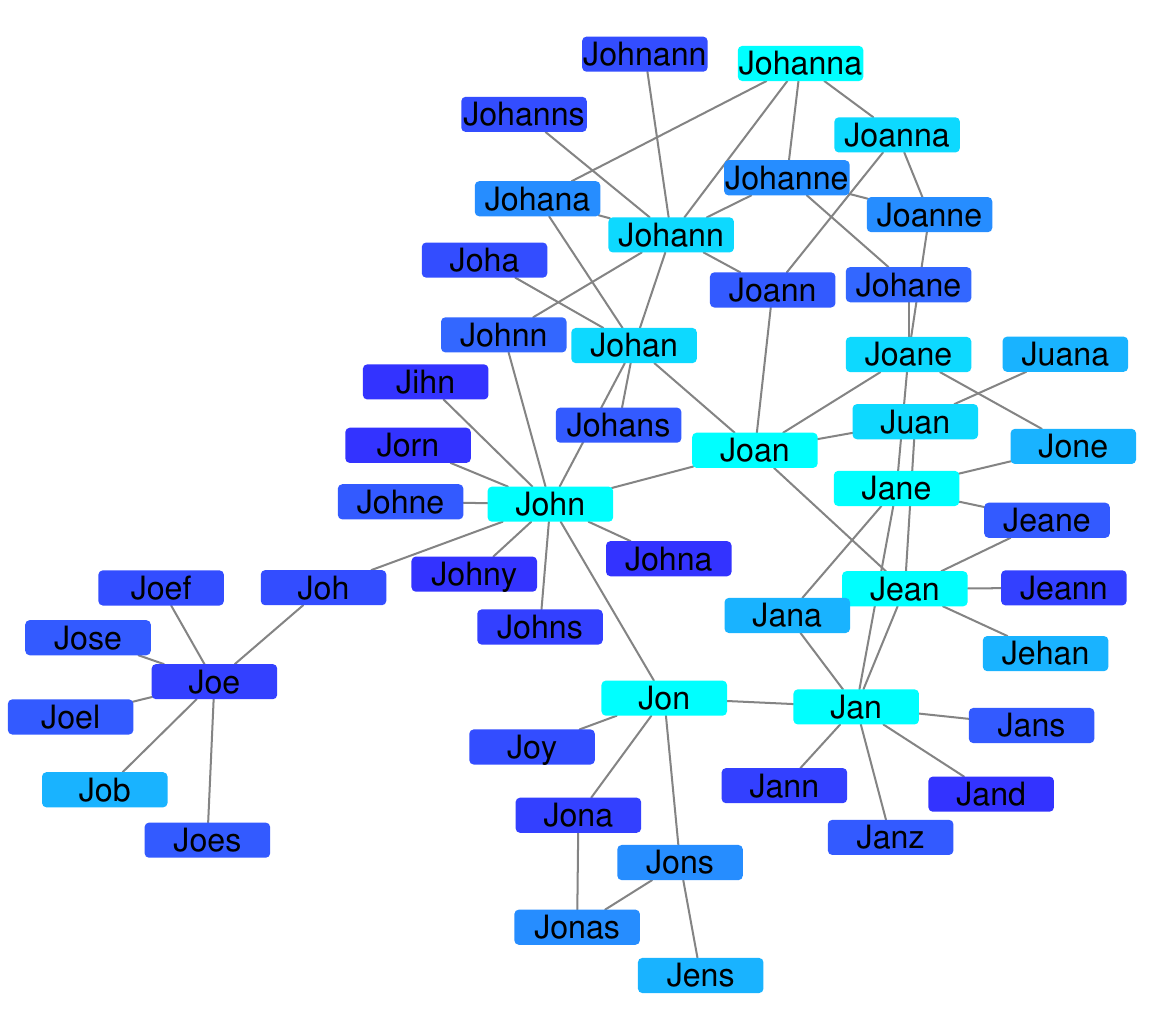}
\caption{A network of synonyms for the name John constructed based on digitized family trees. 
The starting point of this network is John. 
The colors depict the century in which the name first appears.
The centuries range from the 11th century (turquoise) to the 19th century (dark blue).}
\label{fig:name_network_example}
\end{figure}

The main reason for the poor results provided by well known Web search engines stems from the name-containing queries themselves. 
As opposed to a simple word with one correct spelling, there can be many legitimate spelling variations for a personal name~\cite{christen2006comparison}. 
Furthermore, forenames and surnames sometimes change over time due to marriage, religious conversion (e.g., from Cassius Clay Jr. to Muhammad Ali), and gender reassignment (e.g., from Yaron Cohen to Dana International). 
In addition, names are heavily influenced by a person's cultural background~\cite{smalheiser2009author}. 
For instance, the English forename John has several variations in other languages: Jean (French), Giovanni (Italian), Johannes (German and Latin), Jo\~ao (Portuguese), and Juan (Spanish)~\cite{familyeducation} (see Figure~\ref{fig:name_network_example}). 
%This is further complicated by the fact that some people are widely known by their nicknames, and detecting aliases for names is a quite challenging task. 
%For example, Lionel Messi, the famous football player, is called ``La Pulga'' (the flea) and ``Messiah.''
Therefore, matching personal names (forenames and surnames) is a more challenging task for search engines than matching general text.%~\cite{borgman1992getty}.

The name matching problem is well known and has been explored in many research fields, including statistics, databases, record linkage, and artificial intelligence~\cite{cohen2003comparison}.
Today, most techniques for related name retrieval are based on pattern matching, phonetic encoding, or a combination of these two approaches~\cite{christen2006comparison}.
However, despite the research performed in various domains, the retrieval of related names leads to poor results~\cite{friedman1992tolerating}.

In addition to searching for names online, genealogical research is another online activity that has become popular given the increased availability of digitized genealogical documents and access to the Internet around the world~\cite{heinlen2007genealogy}.
In response to the growing interest in genealogy, several online companies specializing in genealogy, such as MyHeritage\footnote{https://myheritage.com/} %~\cite{myheritage}% 
and WikiTree,\footnote{https://www.wikitree.com/} %~\cite{wikitree}%
have emerged.
Based on personal data provided by users, these companies construct personal digitized family trees.
Over time, the many individually constructed family trees merge into a single enormous forest by utilizing the crowd's wisdom and entity matching~\cite{kaplanis2018quantitative}.  

In this paper, we propose an innovative algorithm that addresses the synonym suggestion problem. 
Our novel algorithm utilizes ancestral data collected from digitized family trees, combined with graph algorithms and genealogy (the study of families, family history, and family lineage).%~\cite{yakel2004seeking}.
In contrast to previous approaches that retrieve synonyms based on the same encoded representation or pattern%~\cite{holmes2002improving,uzzaman2004bangla}
, we propose a general approach that suggests personal names (forenames and surnames) based on the construction and analysis of digitized family trees assembled by millions of people in a collaborative effort to trace their past.
Namely, we collected data from a genealogy website to construct a large weighted graph of names which contains information about how they have evolved over the centuries (see Figure~\ref{fig:name_network_example} and Section~\ref{sec:methods}).
Suggesting synonyms using the \emph{GRAph based on names derived from digitized Family Trees (GRAFT)} algorithm provides significantly superior performance compared to other existing algorithms that focus on encoding or the detection of specific patterns.
For example, the average precision at one (precision@1) obtained by \graft is \textit{three times higher} than that of the well-known Soundex algorithm (0.294, as opposed to 0.093) (see Section~\ref{sec:results} and  Table~\ref{tab:top_10_performance_wikitree_forenames}).
This means that the number of correct synonyms suggested using \graft is significantly higher compared to phonetic encoding algorithms, such as Soundex, which suggests synonyms based on similar sounds.   

The remainder of this paper is organized as follows:
In Section~\ref{sec:related_work}, we provide a brief overview of research focused on issues similar to those addressed in this study.
Section~\ref{sec:methods} describes the proposed method for suggesting synonyms, which is based on the construction and analysis of digitized family trees. 
In Section~\ref{sec:data}, we provide a detailed description of the datasets used in this study, and we review our experimental setup in Section~\ref{sec:experimental_setup}. 
The performance of the \graft algorithm, as well as other phonetic and string similarity algorithms, on the task of suggesting synonyms, is presented in Section~\ref{sec:results}.
In Section~\ref{sec:discussion}, we discuss the results obtained, and Section~\ref{sec:conclusion} presents our conclusions and directions for future research.

\section{Background}
\label{sec:related_work}
In the following subsections, we provide an overview of related work and the %necessary% 
background for this study.
More specifically, in Section~\ref{sec:digitized_family_tree_usages}, we review the topic of digitized family trees and their uses.
Next, in Section~\ref{sec:graph_based_approaches}, we present existing graph-based approaches.   
Then, in Sections~\ref{sec:string_similarity_metrics} and \ref{sec:phonetic_encoding_algorithms}, we provide a brief overview of a few well-known string similarity and phonetic algorithms, whose performance we compare to the  \emph{GRAFT} algorithm's performance later in the paper.
Finally, in Section~\ref{sec:related_name_suggestion}, we review prior studies that suggested synonyms based on a given name.

\subsection{Digitized Family Tree and Their Uses}
\label{sec:digitized_family_tree_usages}
Three decades ago, the creation and use of digitized family trees was very limited due to their reliance on the domestic data repositories of churches and record offices~\cite{albright2008utah}.
The two main reasons for such limited use were the lack of comprehensive and accurate genealogical %information% 
data among large populations~\cite{thorsson2003systematic} and the extensive effort needed to digitize and organize the genealogical records~\cite{albright2006computerized}.

However, over the last two decades, there has been an impressive increase in the digitization of genealogical documents and the availability of such documents online.
Today, many universities, libraries, and public institutions digitize these documents to preserve this valuable information and provide open access to them.
The phenomenon of open access to genealogical documents, together with people's growing interest and curiosity regarding their origins~\cite{smolenyak2004trace}, has contributed to the popularity of online genealogical research~\cite{heinlen2007genealogy}.

Today's family trees visually present a person's ancestry simply and conveniently; in most cases, the tree's structure depicts a mathematical graph that attempts to capture natural processes, such as marriage and parenthood~\cite{kaplanis2018quantitative}. 
This structure, which is based on one's ancestors, is an important and useful tool for observing a family's evolution over generations by presenting the relationships between family members~\cite{kingman1982genealogy}.
Furthermore, the valuable information captured by these trees can be utilized in a wide range of research domains. 
Currently, the main research domain that utilizes family trees is genetics, which leverages genotype data from relatives%~\cite{kong2008detection}%
, analyzes parent-of-origin effects%~\cite{kong2009parental}%
, estimates heritability%~\cite{ober2001genetic}%
, and studies disease prevention~\cite{valdez2010family}.
Beyond genetics, family trees have played a major role in domains, such as human evolution%~\cite{lahdenpera2004fitness}%
, anthropology%~\cite{helgason2008association}%
, economics%~\cite{modalsli2016multigenerational}%
, and even behavior analysis over generations~\cite{modalsli2016multigenerational}.
%~\cite{mann1985reliability}. 
Inspired by the convenience and simplicity of presenting the evolution of families over generations, researchers have also utilized this concept to analyze the evolution of myosin protein%~\cite{hodge2000myosin}% 
and cancer~\cite{chung2002molecular}.

In recent years, data science researchers have analyzed large genealogical datasets; in 2015, Fire and Elovici~\cite{fire2015data} applied machine learning algorithms on a genealogical dataset containing data from over %a million% 
1M individuals to discover features that affect individuals’ lifespans over time. 
In 2018, Kaplanis et al.~\cite{kaplanis2018quantitative} obtained a genealogical dataset from Gini.com, which consists of over 86M publicity profiles. 
After an extensive cleaning process, they constructed family trees in which the largest pedigree consisted of 13M people.
They analyzed these family trees and provided insights into population genetic theories. 
In the same year, Charpentiera and Gallic~\cite{charpentier2018internal} used the digitized family trees of 2.5M individuals collected from the Geneanet website to study internal migration in France in the 19th century.

In addition to researchers who used family trees for network evolution analysis, the desire of people to learn about their origins created a new market for online companies that specialize in genealogy~\cite{heinlen2007genealogy}.  
Examples of these companies are Ancestry,\footnote{https://www.ancestry.com/%~\cite{ancestry}
} %FamilySearch,\footnote{https://www.familysearch.org/}%~\cite{familysearch},% 
MyHeritage, %~\cite{myheritage}%
and WikiTree. %~\cite{wikitree}. 
These companies encourage genealogy enthusiasts to upload their family trees by creating profiles for each family member~\cite{kaplanis2018quantitative}.
In many cases, the profile includes basic information such as first and last name, nickname, demographic information, birth and death date, and a photo.
Currently, the popularity of these companies has grown, and each company now boasts millions of customers worldwide.%~\cite{wikitree,ancestry_about,myheritage_about,familysearch_about}. 
The relative advantage of these companies is the scanning operations they perform to detect similar profiles using entity-matching metrics. 
When detecting similar profiles, the websites encourage customers to merge two given profiles into a single profile~\cite{kaplanis2018quantitative}, connecting separate digitized family trees into a larger family tree.
These larger trees provide additional information about the user's ancestors beyond his or her knowledge.

\subsection{Graph-Based Approaches}
\label{sec:graph_based_approaches}
Over the past 20 years, technological development has accelerated thanks to the Internet.
However, the performance in some domains associated with Web search and document retrieval has not advanced similarly, due to the need to search for names, which is challenging due to name ambiguity in which many people %or objects% 
share identical names~\cite{fan2011graph}. 

As a result, many studies have addressed the problem of personal name ambiguity by utilizing graph-based approaches that have been found useful in word sense disambiguation research~\cite{guo2011graph}.
In most of these approaches, two main activities are performed to address the issue of name ambiguity: 
1) the construction of graphs based on the available data, and 2) the application of unsupervised machine learning algorithms (e.g., clustering) to detect similar entities.
McRae-Spencer and Shadbolt~\cite{mcrae2006also} constructed self-citation and co-authorship graphs, and evaluated their approach on large-scale citation networks. 
Fan et al.~\cite{fan2011graph} and Shin et al.~\cite{shin2014author} utilized just the attribute of co-authorship among authors to solve the same problem. 
Bollegala et al.~\cite{bollegala2008co} constructed word co-occurrence graphs to represent the mutual relations between words that appear in anchor texts.
Jiang et al.~\cite{jiang2009grape} utilized tag information to construct tag-based graphs. 
Applying clustering on the graph constructed, they detected people instances.
Similarly, Tang et al.~\cite{tang2011bipartite} utilized the social network snippet of a specific user for the construction of a bipartite graph; using a clustering algorithm, they identified personal entities. 
Guo et al.~\cite{guo2011graph} and Han et al.~\cite{han2011collective} used graph-based entity linking to help detect name mentions in text with their referent entities in a knowledge base.
Smirnova et al.~\cite{smirnova2010using} utilized the link relationships among Web pages for name resolution in Web search results.
By constructing a Web graph and applying clustering, they successfully performed a person name resolution using the Web structure as the only input information.

\subsection{String Similarity Algorithms}
\label{sec:string_similarity_metrics}
In this study, we utilize digitized family trees by connecting family members who share a similar name to their ancestors. 
The condition for detecting similar names is determined by well-known string similarity functions.
In the past, these functions have usually been used to match individuals or families between samples and censuses for tasks like measuring the coverage of a decennial census or combining two databases, such as tax information and population surveys~\cite{casanova2007database}.
Such functions attempt to determine the similarity of two strings by measuring the ``distance'' between the two strings.
Two strings that are found similar by the functions are considered related. 
In this study, we use the following string similarity functions:

\textbf{Damerau-Levenshtein Distance (DLD).} 
The Damerau Levenshtein distance was developed in 1964 by Damerau~\cite{damerau1964technique}.
To transform a given word to another, this string algorithm calculates the minimal number of four different types of editing operations: insertion, deletion, permutation, and replacement.

\textbf{Edit Distance (ED).}
The edit distance, also known as the \emph{Levenshtein distance}, was developed two years later in 1966 by Levenshtein~\cite{levenshtein1966binary}.
This similarity string algorithm calculates the minimal number of operations (insertions, deletions, and substitutions of a single character) required to transform one word into an other~\cite{levenshtein1966binary}. 
For example, the \emph{edit distance} between the names John and Johan is one.

\textbf{Jaro-Winkler Distance.}
This string distance metric, developed in 1995 by Jaro and Winkler~\cite{jaro1995probabilistic}, was intended primarily for short strings like personal surnames~\cite{cohen2003comparison}.
It is based on the number and order of the common characters between two given strings~\cite{cohen2003comparison}.
The lower the Jaro Winkler distance for two strings is, the more similar the strings are.
This is normalized such that zero means an exact match, and one means there is no similarity.
In this study, we used the Jaro Winkler similarity metric, which is the inversion of the distance metric described above.

\subsection{String Similarity Join}
\label{sec:string_similarity_joins}

In contrast to string similarity measures which estimate the similarity between two strings for approximate string matching or comparison, string similarity join (SSJ) searches for similar string pairs between two string sets~\cite{wang2011fast}.
SSJ usually serves as an essential operation in many applications in areas, such as data integration and deduplication.

In 2011, Wang et al.~\cite{wang2011fast} proposed fuzzy token matching based similarity, which extends token-based similarity functions, such as Jaccard and cosine similarity, by enabling fuzzy matching between two tokens.
%In addition, t
They presented a novel signature scheme for token sets and developed pruning techniques to improve performance. 
Their evaluation showed that their approach achieved %high efficiency and result quality.
higher performance.

In 2013, Lu et al.~\cite{lu2013string} presented two expansion-based methods to quantify the similarity of strings and I-tree, an index structure titled for performing similarity join.
In addition, they developed an estimator for selecting signatures online to increase the efficiency of signature filters in join algorithms.
They evaluated their methods on three datasets and showed their advantages over state-of-the-art methods.

In 2018, Tao et al.~\cite{Tao2017ApproximateSJ} studied approximate string joins with
abbreviations (ASJA). 
The authors proposed an innovative similarity measure that estimates the similarity between two strings by taking abbreviations into account. 
They presented PTIME, a join algorithm which uses filter verification to reduce time complexity, as well as an unsupervised approach for learning a dictionary of abbreviation rules from input strings based on the LCS assumptions. 
Evaluation was performed % their methods %
on four real-world datasets and showed their effectiveness over state-of-the-art approaches.

\subsection{Phonetic Encoding Algorithms}
\label{sec:phonetic_encoding_algorithms}
We compare \emph{GRAFT}’s performance to that of the phonetic encoding algorithm family.
These algorithms transform a given word into code based on the way the word is pronounced. 
They are commonly used for spelling suggestion%~\cite{uzzaman2004bangla}%
, entity matching%~\cite{cohen2003comparison,peled2013entity}%
, and searching for names on websites~\cite{khan2017application} or in databases.%~\cite{patman2001soundex}.
In this paper, we evaluate the Soundex, Metaphone, Double Metaphone, the New York State Identification and Intelligence System Phonetic Code (NYSIIS), and the match rating approach (MRA).

\textbf{Soundex.} Devised over a century ago by Russel and O’Dell, the Soundex algorithm is one of the first phonetic encoding techniques~\cite{hall1980approximate}.
Given a name, it provides a code that reflects how the name sounds when spoken.
It keeps the first letter in a given name and reduces all of the remaining letters into a code consisting of one letter and three digits.
Vowels and the letters \textit{h} and \textit{y} are converted to zero. 
The letters \textit{b}, \textit{f}, \textit{p}, and \textit{v} are converted to one.
The letters \textit{c}, \textit{g}, \textit{j}, \textit{k}, \textit{q}, \textit{s}, \textit{x}, and \textit{z} are converted to two. 
The letters \textit{d} and \textit{t} are converted to three, whereas \textit{m} and \textit{n} are converted to five.
The letter \textit{l} is converted to four, whereas \textit{r} is converted to six.
The final code includes the original first letter and three numbers.
Codes that are generated based on longer names are cut off, whereas shorter codes are extended with zeros.
For example, the Soundex code for the name Robert is R163.  

\textbf{Metaphone.} The Metaphone algorithm was developed in 1990 by Philips~\cite{philips1990hanging}.
It is an improvement over Soundex, because the words are encoded into a more general representation containing only alphabetic characters instead of numbers like Soundex so that they can be combined into a group despite minor differences.%~\cite{binstock1995practical}.
This algorithm assumes English phonetics and works equally well for forenames and surnames.%~\cite{pimpalkhute2014phonetic}.
It is widely used in spell checkers, search interfaces, genealogy websites, etc.%~\cite{khan2017application}.
As an example, the Metaphone code for the forename Robert is RBRT.

\textbf{Double Metaphone.} The Double Metaphone algorithm was developed almost two decades ago by Philips~\cite{philips2000double}.
The Double Metaphone is a variation of the Metaphone algorithm which generates a code that consists solely of letters.
As opposed to the previous two algorithms, the Double Metaphone also attempts to encode non-English words (European and Asian names).
Moreover, unlike all other phonetic algorithms, it suggests two phonetic codes.
As an example, the Double Metaphone codes for the forename Jean are JN and AN. 

\textbf{NYSIIS.} The New York State Identification Intelligence System (NYSIIS) also provides a code that consists solely of alphabetic letters~\cite{borgman1992getty}. 
It preserves the vowels' positions in the given name by converting all of the vowels to the letter `A.'%~\cite{de1986guth}. 
The NYSIIS code for the forename Robert is RABAD.

\textbf{Match Rating Approach (MRA).}
This phonetic encoding algorithm was developed by Gwendolyn Moore in 1977~\cite{moore1977accessing}. 
The algorithm includes a small set of encoding rules, as well as a more lengthy set of comparison rules. 
The MRA code for the forename Robert is RBRT.

 \subsection{Related Name Suggestion Algorithms}
 \label{sec:related_name_suggestion}
In 1996, Pfeifer et al.~\cite{pfeifer1996retrieval} examined the differences in the performance of a few known phonetic similarity measures and exact match metrics for the task of improving the retrieval of names.
For evaluation, Pfeifer et al. manually collected surnames from a few sources, such as the TREC collection%~\cite{harman1992overview}%
, the CACM collection from the
SMART system%~\cite{buckley1985implementation}%
, the phonebook of the University of Dortmund in Germany, and author names from a local bibliographic database.
At the end of this process, all of the surnames were combined into one large dataset entitled COMPLETE, containing approximately 14,000 names.
Afterward, they defined the queries for this dataset as follows: 
First, they chose 90 names randomly from the COMPLETE dataset.
Second, for each of the 90 queries selected, they manually determined the relevant names.
They showed that an information system based on phonetic similarity measures, such as Soundex, and variations of phonetic algorithms outperformed exact match search metrics in the task of searching related names.

In 2010, Bollegala et al.~\cite{bollegala2010automatic} presented a method for extracting aliases for a given name based on the Web.
For example, the alias of ``the fresh prince'' is Will Smith. 
They proposed a lexical pattern-based approach for extracting aliases of a given name, using snippets returned by a Web search engine.
Later, they defined numerous ranking scores to evaluate candidate aliases, using three approaches: lexical pattern frequency, word co-occurrences in an anchor text graph, and page counts on the Web.
Their method outperformed numerous baselines, achieving a mean reciprocal rank of 0.67.
%In contrast to personal names, 
Suggesting nicknames is a challenging task, since in many cases they are not necessarily associated with the given name.
Therefore, in this paper, we focused on suggesting synonyms that share ancestral roots, such as Elisabeth and Lisabeth, Samuel and Sam, Sophia and Sophie, Robert and Robbie, etc.

%For example, Lionel Messi, the famous football player, is called ``La Pulga'' (the flea) and ``Messiah.''
%In this paper, nickname suggestion is out of scope.}

In 2019, Foxcroft et al.~\cite{foxcroft2019name2vec} presented Name2Vec, a machine and deep learning method for name embeddings that employs the Doc2Vec methodology, where each surname is viewed as a document, and each letter constructing the name is considered a word.
They performed the task of record linkage by training a few name embedding models on a dataset containing 250,000 surnames and tested their model on 25,000 verified name pairs from Ancestry. 
They used the Ancestry Records dataset as 25,0000 positive samples and another 25,000 random name pairs as negative samples.
The authors concluded that the name embeddings generated can predict whether a pair of names match. 

Along with research aimed at suggesting synonyms, several companies emerged to address the task of finding people based on their names. 
The companies responding to the growing need of Internet users to find people online and the poor results provided by the large search engines\footnote{https://organicweb.com.au/social-media/people-search-pipl-wink-peekyou/} %~\cite{organicweb}
include Pipl %~\cite{pipl}%
, which utilizes names to search for the real person behind online identities%~\cite{pipl_about_us}% 
and ZoomInfo, %~\cite{zoominfo}%
which provides information about people that is company or organizational-oriented.
According to ZoomInfo,\footnote{https://www.zoominfo.com/business/about} %~\cite{zoominfo_about_us}%
their data includes 67M email addresses and 20M company profiles.

Other similar online services that are free are PeekYou,\footnote{https://www.peekyou.com/} %~\cite{peekyou}%, 
a people search service that collects and combines content from online social networks, news sources, and blogs to help retrieve the online identity of American users, and True People Search,\footnote{https://www.truepeoplesearch.com/} %~\cite{truepeoplesearch},% 
which helps find people by name, phone number, or address.
Websites, such as TruthFinder\footnote{https://www.truthfinder.com/} %~\cite{truthfinder}% 
%and BeenVerified,\footnote{https://www.beenverified.com/}%~\cite{beenverified}%, 
provide background checking services for people. 
These services can help reconnect Americans with their friends and relatives, as well as provide a way to look up criminal records online.

\section{Methods}
\label{sec:methods}

In this paper, we propose \emph{GRAFT}, a novel algorithm for improving the suggestion of synonyms associated with a given name.
Our pioneering algorithm is based on the construction and analysis of digitized family trees, combined with network science.
By constructing digitized family trees, we utilize the valuable ancestral information that exists in these family trees to detect family members who share a similar name.
Afterward, by connecting names that many family members have preserved over generations, we construct a graph based on names that reflects the evolution of names over generations (see Figure~\ref{fig:name_network_example}).
Then, we search for the given name in the graph and select candidates to be suggested as synonyms according to a general ordering function that can consider a few parameters, such as the network's structure and the string and phonetic similarity between the given name and the candidate (see Section~\ref{sec:suggesting_related_names_based_on_name_based_networks}).

\subsection{Suggesting Synonyms Based on a Graph of Names}
\label{sec:suggesting_related_names_based_on_name_based_networks}

\begin{figure*}[h!]
\centering
\includegraphics[scale=0.6]{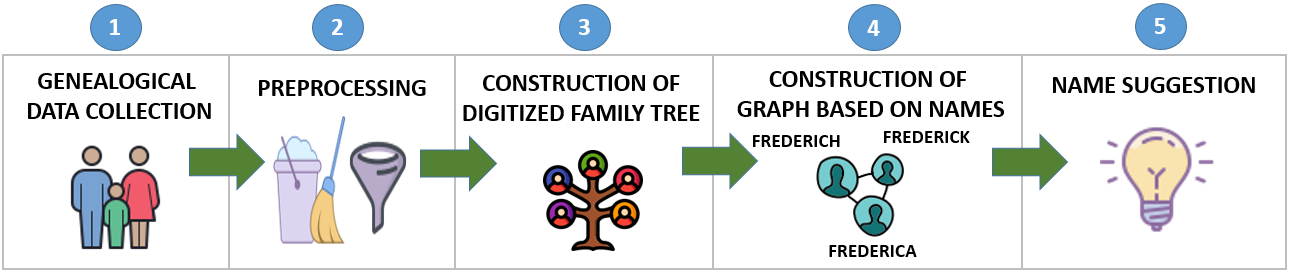}
\caption{Overview of \emph{GRAFT}'s phases for discovering and suggesting synonyms.}
\label{fig:proposed_method}
\end{figure*}

The proposed method consists of five main phases: data collection, preprocessing, construction of a digitized family tree, construction of a graph based on names, and name suggestion (see Figure~\ref{fig:proposed_method}). 

\begin{enumerate}
\item \textbf{Genealogical Data Collection.} 
The proposed \graft algorithm utilizes the inherent ``wisdom'' that exists in digitized family trees.
Therefore, in the first phase, we use a genealogical dataset which includes valuable information regarding people and their ancestors, such as forenames and surnames, nicknames, parents' names, and more.  

\item \textbf{Preprocessing.} 
In this phase, we clean short abbreviations and prefixes associated with forenames and surnames.
For example, a person named ``Aaron T Jones,'' is changed to ``Aaron Jones'' because the ``T'' character is an abbreviation of an unknown middle name.
We remove all of the names with less than two characters in order to avoid abbreviations. 
Similarly, very common prefixes, which mainly serve as prepositions meaning ``of'' or ``from'' in different languages (e.g., van in the Dutch and Afrikaans languages, de in Latin, Da in Italian), or ``the'' (e.g., Der in German, La in Italian, French, and Spanish, Le in French) or ``of the'' (e.g., Das and Dos in Portuguese, Dele in French and Spanish, Du in French) are also removed.
% Similarly, very common prefixes mostly related to surnames, such as van, de, la, dos should also be removed. 
% Van is a common prefix which in many cases refers to a quite distance ancestor's place of origin or residence, as well as a preposition in the Dutch and Afrikaans languages, meaning ``of'' or ``from.''
% For example, Rembrandt van Rijn meaning ``from the Rhine'' (one of the major European rivers). 
%the given names, such as the forenames and surnames of people who use short abbreviations.
In addition, we remove English honorific titles, such as Mr., Dr., Jr., etc.\footnote{While short names are widely used in public, using these names to construct the digitized family trees will harm our analysis of the evolution of names.}

\item \textbf{Constructing Digitized Family Trees.} 
Using the cleaned genealogical dataset, we construct digitized family trees, forming a huge graph, by linking child and parent profiles to each other. 
More specifically, we construct a direct graph $G_T := <V_T, E_T>$, where $V_T$ is a set of profiles in the cleaned genealogical dataset, and $E_T$ is a set of directed links between profiles; each link, $e:=(u, v) \in E_T$, connects two profiles $u,v \in V_T$, where $u$ is a parent of $v$. %(see Figure~\ref{fig:family_tree_construction}). 
At the end of this step, a large graph with millions of vertices and links is created.

% \begin{figure}[h!]
% \centering
% \includegraphics[scale=0.5]{construct_family_tree.JPG}
% \caption{A demonstration of a digitized family tree constructed based on given genealogical records.}
% \label{fig:family_tree_construction}
% \end{figure}

\item \textbf{Constructing a Graph Based on Names.} 
By using $G_T$, we create a new weighted graph in which each vertex is a name, each link connects names of a parent and his/her children, and each link's weight is the number of times links between two exact names exists in $G_T$. 
To reduce the size of the graph, we only establish links between two vertices, where the ``distance'' between their names is small.
Namely, we create a name-based graph
$G_N^{i,j} := <V_N, E_N^{i,j}>$, where $V_N$ is a set of vertices defined as follows: $V_N := \{n| \exists v \in V_T \mbox{ and } v_{name} = n \}$, and $v_{name}$ is defined as the name of a profile $v \in V_T$ (can be a forename or surname). 
Additionally, we define $E_N^{i,j}$ to be the following set of links $e:=(n,m,w) \in E_N$, where $n,m \in V_N$ and  $\exists (u,v) \in E_T$,
$u_{name} = n, v_{name} = m$, and  $edit-distance(n,m)\in [i,j]$.
Moreover, we define $w > 0$ to be equal to the following: $w :=|\{(u,v) \in E_T| u_{name}=n \mbox{ and } v_{name} = m \}|$, i.e., the number of times two exact names exist in $G_T$. 
Lastly, we remove links between two vertices if their names are too far apart.

\item \textbf{Name Suggestion.} 
By using $G_N^{i,j}$, we suggest synonyms as follows (see Figure~\ref{fig:suggesting_name_phase}):
given the name $n$, we search for $n\in V_N$.
In cases in which the given name does not exist in $V_N$, no synonyms are returned.
When $\exists  n \in V_n$, we search for potential candidate synonyms using the following steps: 
First, we traverse $G_N^{i,j}$ using the breadth-first search (BFS) algorithm starting from $n$.
This means that in the first iteration, we pass all of the neighbors that are directly connected to $n$. 
Next, in the second iteration, we pass the neighbors of $n$'s neighbors, and so forth.
In order to accelerate the search procedure, we recommend constructing indexes for each name and its neighbors prior to the traversing phase.
After passing all of the vertices reachable from $n$ (defined as $R_n \subseteq V_N$), we provide a score to each reachable vertex $r\in R_n$, according to the predefined name similarity scoring function $f: (n,r) \rightarrow \mathbb{R}^{+}$, which measures how similar each reachable vertex $r\in R_n$ is to $n$. 
The score provided is generated by one of the following four ordering functions: $NetED$, $Net^2ED$, $EDofDMphone$, and $NetEDofDMphoneED$  (see Table~\ref{tab:ordering_functions_description}), %($Similarity_i, i=1..4$): 
where $n_1$ and $n_2$ are names, $SP(n_1, n_2)$ is a function that retrieves the shortest path from the starting vertex to the goal vertex in $G_N^{1,3}$, $ED := ED(w_1, w_2)$ is a function that returns the minimal number of editing operations required to transform from word $w_1$ into $w_2$ (see Section~\ref{sec:string_similarity_metrics}), and $DM : = DoubleMetaphone$ is a function that returns the phonetic sound code of a given name.

The motivation behind the first function $NetED$ is to take the similarity between the names in two dimensions into account: first, in the sense that the names as strings are similar, and second, in the sense that the names' vertices are also near each other in the given $G_N^{1,3}$.
The second function is $Net^2ED$ which is similar to $NetED$ however it prioritizes the proximity of the names in the graph.
In contrast to $NetED$ and $Net^2ED$ which combine the string similarity and network structure, the third function $EDofDMphone$ focuses on the performance of phonetic algorithms. 
For the phonetic algorithm, we chose Double Metaphone, because this algorithm improves Soundex and Metaphone, and it provides both a primary and secondary code for a name, a mechanism that can help find different synonyms.
The fourth function is $NetEDofDMphoneED$ which considers all of the factors that can help in name suggestion: name and phonetic similarity, and network structure.

\begin{table*}[t]
  \centering
  \caption{Ordering Function Description}
  \begin{tabular}{ccc}
  Ordering Function & Type & Definition \\ [0.5ex] 
 \hline\hline
$NetED$ & Net, String Sim & $NetED(n_1, n_2) = \frac{1}{SP(n_1, n_2) \cdot {ED}(n_1, n_2)}$  \\
\hline
$Net^2ED$ & Net, String Sim & $Net^2ED(n_1, n_2) = \frac{1}{(SP(n_1, n_2))^2 \cdot {ED}(n_1, n_2)}$  \\
\hline
$EDofDMphone$ & Phonetic, String Sim & $EDofDMphone(n_1, n_2) = \frac{1}{Min(ED(DM(n_1), DM(n_2)))}$ \\
\hline
$NetEDofDMphoneED$ & Net, Phonetic, String Sim & $NetEDofDMphoneED(n_1, n_2) = \frac{1}{SP(n_1, n_2) \cdot {ED}(n_1, n_2) \cdot Min(ED(DM(n_1), DM(n_2)))}$  \\
\hline
  \end{tabular}
  \label{tab:ordering_functions_description}
\end{table*}

% \begin{enumerate}
%         \item $NetED(n_1, n_2) = \frac{1}{SP(n_1, n_2) \cdot {ED}(n_1, n_2)}$
%         \item $Net^2ED(n_1, n_2) = \frac{1}{(SP(n_1, n_2))^2 \cdot {ED}(n_1, n_2)}$
%         \item $EDofDMphone(n_1, n_2) = \frac{1}{Min(ED(DM(n_1), DM(n_2)))}$
%         \item $NetEDofDMphoneED(n_1, n_2) = \frac{1}{SP(n_1, n_2) \cdot {ED}(n_1, n_2) \cdot Min(ED(DM(n_1), DM(n_2)))}$ 
% \end{enumerate}

%Next, after obtaining the potential candidates, we measured the similarity between the given name and each of the candidates by ranking the names retrieved according to one of the proposed ordering functions ($Similarity_i, i=1..4$).
For demonstration, assume a search for the forename Robert, in the name-based graph. 
After detecting this name in the graph, we traverse the graph to collect the following candidate names: Rob and Reuben. 
Both were located at a depth of one from the given name Robert. 
In the next phase of this example, we apply $NetED$ on the given name and its candidates. 
For example, for the name Robert, we calculate the following:

\begin{equation}
\begin{aligned}
&NetED(Robert, Rob) =  \\ 
&=\frac{1}{SP(Robert, Rob) \cdot {ED}(Robert, Rob)} = \\
&= \frac{1}{1 \cdot 3} = \frac{1}{3} \\
\end{aligned}
\end{equation}

\begin{equation}
% \vspace{-2cm}
\begin{aligned}
&NetED(Robert, Reuben) =  \\ 
&=\frac{1}{SP(Robert, Reuben) \cdot {ED}(Robert, Reuben)} = \\
&= \frac{1}{1 \cdot 4} = \frac{1}{4} \\
\end{aligned}
\end{equation}

%   \item $NetED(Robert, Rob) = \frac{1}{SP(Robert, Rob) \cdot {ED}(Robert, Rob)}  = \frac{1}{1 \cdot 3} = \frac{1}{3} $
        
% \item $NetED(Robert, Reuben) = \frac{1}{SP(Robert, Reuben) \cdot {ED}(Robert, Reuben)}  = \frac{1}{1 \cdot 4} = \frac{1}{4} $

Lastly, we sort all of the vertices $R_n$ and suggest the top-$k$ reachable vertices in $R_n$ that received the highest $f$ scores as synonyms.
%Then, we sort the candidates according to the selected ordering function.
Therefore, according to this example, we retrieve Rob, which is followed by Reuben.

\begin{figure*}[h!]
\centering
\includegraphics[scale=0.5]{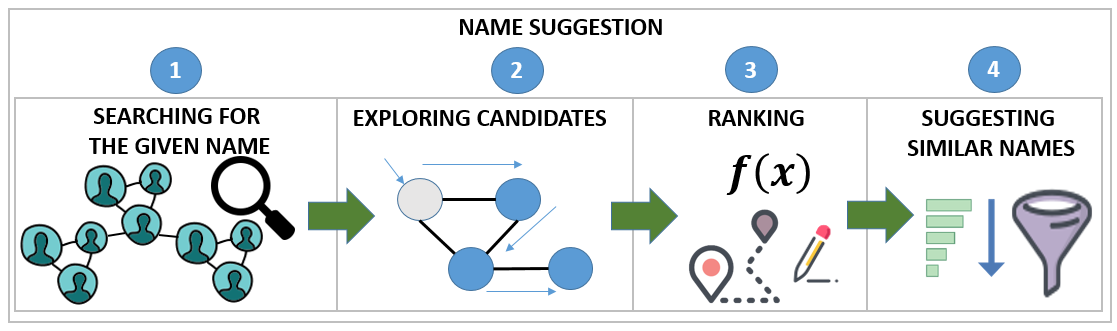}
\caption{Name suggestion steps.}
\label{fig:suggesting_name_phase}
\end{figure*}

\end{enumerate}

\subsection{Suggesting Synonyms Based on Hybrid \emph{GRAFT}}
\label{sec:hybrid_approach}

One of the strengths and unique qualities of the \graft algorithm, as well as one of its drawbacks, is its dependency on the name-based graph for suggesting synonyms.
Of course, the greater the number of names in the genealogical dataset, the greater the likelihood of being suggested synonyms for a specific name.
Still, there may be cases for which \graft would be unable to suggest synonyms for a given name that does not exist in the genealogical dataset.
Therefore, we propose the hybrid \graft (\emph{HGRAFT}) algorithm, a naive and hybrid approach combining \graft and a phonetic encoding algorithm, which can be also be used in cases in which a given name does not exist in the name-based graph constructed; in these cases, synonym suggestions would be suggested based on the phonetic encoding algorithm.
For example, assume that synonym suggestions are needed  for the following three names: Robert, John, and, Felix.
Robert and John exist in the name-based graph, but Felix does not. 
As a result, \hgraft suggests synonyms for Robert and John using the graph and suggests synonyms for Felix using the phonetic encoding algorithm.

\section{Data Description}
\label{sec:data}
In this study, to evaluate our proposed algorithm, we used three datasets: the WikiTree dataset, which includes more than 700K forenames and 500K surnames, the Behind the Name Forenames dataset, which includes 31,552 verified synonyms for 6,274 forenames, the Behind the Name (BtN) Surnames dataset, which includes 22,764 verified synonyms for 3,139 surnames, and the Ancestry Surnames dataset, which consists of 2,500 verified synonyms.
In the following subsections, we describe each dataset:

\subsection{WikiTree Dataset}
\label{sec:wikitree_dataset}

The proposed algorithm relies on a genealogy dataset and inherent knowledge that exists in the dataset's historical records.
Therefore, for evaluation, we used the open genealogical records obtained from the WikiTree website.\footnote{\url{https://www.wikitree.com/wiki/Help:Database_Dumps}} %~\cite{Wikitree_dump}.
Wikitree is an online genealogical website that was founded in 2008 by Chris Whitten.%~\cite{wikitree}.

Free and accessible worldwide, WikiTree’s main goal is to provide an accurate single family tree using genealogical sources.
%The main goal of WikiTree is to provide an accurate single family tree using genealogical sources, a fact that makes genealogy free and accessible worldwide.
As of January 2020, WikiTree had over 760,715 registered users and maintained over 25 million ancestral profiles.%~\cite{wikitree}.
Many of these profiles contain specific details about each individual, such as full name, nickname, gender, birth and death dates, children's profiles, etc.
The massive WikiTree dump we worked with includes more than 16 million profiles and over than 715,000 unique forenames.

\subsection{Behind the Name (BtN) Datasets}
\label{sec:behind_the_name_dataset}

In order to evaluate the performance of the proposed algorithm and compare it to other algorithms, we created two datasets for forenames and surnames by combining the information included in the WikiTree dataset with the data in the Behind the Name (BtN) website. %\footnote{https://www.behindthename.com/}%~\cite{behindthename}. 
This website was founded in 1996 by Mike Campbell to study aspects of given names.\footnote{https://www.behindthename.com/info/} %~\cite{behindthename_info}. 
It contains many given names from various cultures and periods, as well as mythological and fictional names.
Currently, it includes 23,751 names.

The creation of this ground truth dataset was performed as follows:
First, we extracted all of the distinct forenames and surnames in the WikiTree dataset that are comprised of more than one letter. %to avoid English honorific titles.  
Among its over 16 million profiles, we extracted 715,484 unique forenames and 537,192 surnames.   
Using the public service application programming interface (API) provided by the Behind the Name website, we collected synonyms for the distinct names.
For a given forename, there were an average of 5.03 synonyms provided, whereas 7.25 synonyms were provided for each surname. 
As an example, for the forename Bob, we collected the following synonyms: Rupert, Robin, Robbie, Bobby, and Robert.\footnote{https://www.behindthename.com/name/bob/related} %~\cite{bob_behindthename}}.
For the forename Elisabeth, we retrieved Eli, Elisa, Ella, Elsa, Lisa, Liz, and so on.\footnote{https://www.behindthename.com/name/Elisabeth/related} %~\cite{elisabeth_behindthename}.
For the surname Abrahams, we retrieved Abram, Abramson, Abrams, Avraham, Abrahamsen, Ebrahimi, Brams, etc.\footnote{https://surnames.behindthename.com/name/abrahams}
In total, 31,552 synonyms were retrieved for the 6,274 distinct forenames and 22,764 synonyms for 3,139 distinct surnames.
The forenames that provided the maximal number of synonyms were Ina, Nina, and Jan with 111, 105, and 81 synonyms respectively.
The surnames that provided the maximal number of synonyms were Jackson, Johnson, and Jansen with 59, 58, and 58 synonyms, respectively.

\subsection{Ancestry Surnames Dataset}
\label{sec:ancestry_surnames_dataset}

This dataset was collected by Ancestry in 2014 and published by Sukharev et al.~\cite{sukharev2014parallel}. 
Two datasets from that study are utilized: the Records, and Surnames datasets. 
The Records dataset consists of 25K surname pairs, where the second name in each pair serves as a synonym for the first name. 
The second dataset is the Surnames dataset, which contains 250K of the most commonly occurring surnames in Ancestry's overall database.

\section{Experimental Setup}
\label{sec:experimental_setup}

\subsection{Setting Experimental Parameters}
In this study, we aimed at answering three research questions using the experiments described below: 
1) \textit{Is the proposed \graft algorithm useful for the task of suggesting synonyms?} 
2) \textit{How does \emph{GRAFT}'s performance compare to other well-known algorithms, such as phonetic encoding, string similarity and machine and deep learning algorithms?}
3) \textit{Can the proposed algorithm's performance be improved by suggesting synonyms based not only on parents but also grandparents or great-grandparents?} 
(4) \textit{How does the edit distance range affect the size of the name-based graphs constructed?}

\subsubsection{GRAFT Validation}
\label{sec:graft_validation}

To evaluate the proposed \graft algorithm, we executed the following large-scale experiment on each of the datasets examined.
For clarity, we describe the experiment executed on the WikiTree dataset (see Section~\ref{sec:wikitree_dataset}). 
%First, as a data source, we used the WikiTree dataset 
As described above, in the preprocessing phase, we cleaned the forenames by removing short abbreviations and common prefixes that were less than two characters (see Section~\ref{sec:methods}).
Next, we constructed digitized family trees as a large-scale graph, $G_T$ by linking the WikiTree user profiles with those of their parents.
This huge graph $G_T$ consisted of 197,036 vertices, 3,044,150 links, and 154 connected components.
Then, using $G_T$, we generated an additional new weighted forename graph, $G_N$, whose vertices were forenames, and each link $(n_1, n_2, w)$ connected two forenames $n_1$ and $n_2$, with $w$ equal to the number of links in $G_T$ that connected users with the forename of $n_1$ to their parents with the forename of $n_2$ (see Section~\ref{sec:methods}). 
Then, we generated the $G_N^{1,3}$ graph (consisting of 5,906 vertices, 14,940 links, and  439 connected components) by using $G_N$ and leaving only links between related parent and child forenames with edit distance values ranging from one to three.\footnote{We limited the edit distance values so they were less than or equal to three, because we observed that names with edit distance values greater than three provided less relevant synonym suggestions.}
%We limited the edit distance values so they were less than or equal to three because we observed that names with edit distance values greater than three were quite different from the searched name and found that in the vast majority of the cases the names will not provide relevant synonym suggestions.}

Then, we searched for each of the forenames that share synonyms in the ground truth in $G_N^{1,3}$. 
If the searched name appeared in $G_N^{1,3}$, we traversed the graph using BFS, starting from the given name and collecting its neighbors, up to a depth of two.
If the name did not appear in $G_N^{1,3}$, we moved on to the next name on the list.

Next, after obtaining the potential candidates, we measured the similarity between the given name and each of the candidates by ranking the names retrieved according to one of the proposed ordering functions ($Similarity_i, i=1..4$).
Then, we sorted the candidates according to the score provided by the selected ordering function.

% \bluecomment{It is important to point out that the pipeline described also was carried out for the surnames existing in the WikiTree dataset.} 

\subsubsection{Fine-Tuning \emph{GRAFT}}

One of the major contributions of this study is the suggestion of synonyms by utilizing the name-based graphs derived from digitized family trees. 
The graphs were constructed based on children whose name is a variation of one of their parents' names. 
These graphs should reflect the evolution of names over the centuries.
Famous examples of figures who were named after their parents are George W. Bush, the 43rd president of the United States, who was named after his father, George H. W. Bush, the 41st president; William II, King of England, who was named after his father, William the Conqueror; Henry VIII, King of England, who named after his father, Henry VII of England.

However, there are also many cases in which parents named their children after their grandparents (e.g., Cambyses II who was the second king of the Achaemenid Empire, and the son and successor of Cyrus the Great) and even their great-grandparents (e.g., Prince George of Cambridge, who was named after George VI, King of the United Kingdom, who was the great-grandfather of his father Prince William).
Therefore, we evaluate the strength of two of our algorithm's parameters for improving performance: the edit distance range and the identity of the figures for whom we wish to construct a name-based graph.

\subsection{Evaluation Process}
\label{sec:evaluation}

To analyze and evaluate the performance of the proposed \graft algorithm on the task of name suggestion, we evaluated its performance (see Section~\ref{sec:graft_evaluation}), as well as the performance of other algorithms used for suggesting synonyms, such as phonetic encoding algorithms (see Section~\ref{sec:phonetic_encoding_algorithms_evaluation}), string similarity algorithms (see Section~\ref{sec:string_similarity_algorithms_evaluation}), and the more recently proposed Name2Vec algorithm (see Section~\ref{sec:name2vec_evaluation}).

\subsubsection{Evaluation of \graft}
\label{sec:graft_evaluation}

\textbf{Evaluation on Forenames.}
We searched for each forename that appeared in the ground truth using $G_N^{1,3}$.
If the name appeared in $G_N^{1,3}$, we traversed the graph using BFS, starting from the given name and collecting its neighbors up to a depth of two.
Next, we sorted these neighbors according to the score provided by each ordering function proposed in Section~\ref{sec:graft_validation}.
Then, we evaluated the performance of the top 10 suggestions provided.
The evaluation was performed by differentiating between the suggestions and the correct synonyms in the ground truth.
Coverage was measured by counting the number of distinct names in the ground truth for which \graft was successful in suggesting synonyms.
In addition, we used the performance metrics of accuracy, F1, precision, and recall. 
Concerning the precision measure, for each given name in the ground truth, we calculated the metric of $average\mbox{-}precision@k$ where $k = 1, 2, 3, 5, 10$.
We chose to evaluate the top 10 suggestions, because like any search for results in any search engine, people are still only willing to look at the first page of results~\cite{brin1998anatomy}.

\textbf{Evaluation on Surnames.}
Using the WikiTree dataset, we cleaned the surnames by removing short abbreviations that had less than one character.
Next, we constructed digitized family trees as a large-scale graph, $G_T$ by linking the WikiTree user profiles with those of their parents.
This huge graph $G_T$ consisted of 364,383 vertices, 2,702,441 links, and 8,768 connected components.
Then, using $G_T$, we generated an additional new weighted graph, $G_N$, whose vertices were surnames, and each link $(n_1, n_2, w)$ connected two surnames $n_1$ and $n_2$, with $w$ equal to the number of links in $G_T$ that connected users with the surname of $n_1$ to their parents with the surname of $n_2$ (see Section~\ref{sec:methods}). 
Then, we generated the $G_N^{1,3}$ graph (consisting of 26,557 vertices, 21,038 links, and 8,356 connected components) by using $G_N$ and leaving only links between related parent and child surnames with edit distance values ranging from one to three.
%\footnote{We limited the edit distance values so they were less than or equal to three, because we observed that names with edit distance values greater than three provided less
%relevant synonym suggestions.}
%were quite different from the searched name, and in the vast majority of the cases the names will not provide relevant synonym suggestions.}
Next, we sorted these neighbors according to the score provided by each ordering similarity function proposed.
Then, we evaluated the performance of the top 10 suggestions provided.

In addition, we conducted an additional experiment, in which we constructed a name-based graph, using the names of grandchildren and their grandparents.
In this case, the huge graph $G_T$ consisted of 197,512 vertices, 1,531,590 links, and 854 connected components.
Then, using $G_T$, we generated a surname graph, $G_N$, whose vertices were surnames, and each link $(n_1, n_2, w)$ connected two surnames $n_1$ and $n_2$, with $w$ equal to the number of links in $G_T$ that connected users with the surname of $n_1$ to their grandparents with the surname of $n_2$ (see Section~\ref{sec:methods}). 
Then, we generated the $G_N^{1,3}$ graph (consisting of 17,769 vertices, 15,419 links, and  4,000 connected components) by using $G_N$ and leaving only links between related grandparent and grandchild surnames with edit distance values ranging from one to three.
The evaluation for those experiments was carried out using the three datasets.% Ancestry Records dataset, the same dataset used by Foxcroft et al.

Concerning time complexity, we measured the time it took to clean the data, as well as the time required to construct the digitized family tree and the name-based graphs. 
In addition, we considered the execution time required for each ordering function to suggest synonyms. 
Finally, we summed all of the amounts of time into a single aggregated period.

\subsubsection{Evaluation of \hgraft}
In order to evaluate the naive approach proposed in Section~\ref{sec:hybrid_approach}, we suggested synonyms for the names in the ground truth datasets (BtN Forenames and Surnames, and Ancestry Surnames datasets), as follows:
In cases in which a given name was in the name-based graph, we suggested synonyms according to \emph{GRAFT,} with the   $NetEDofDMphoneED$\footnote{Achieved the best performance among the four proposed similarity functions} similarity function, using a threshold for the edit distance ranging from one to three.
In cases in which a given name is not in the name-based graph, synonyms were suggested using the phonetic encoding algorithm which achieved the best performance (Double Metaphone for forenames and NYSIIS for surnames).
%Next, all the names who share the same phonetic code were retrieved as optional candidates. 
%Then, the top 10 candidates were sorted according to their edit distance from the given name in ascending order.  

Regarding time complexity, we took into account the time spent for suggesting synonyms according to both algorithms, separately, as well as the time it took to suggest synonyms again for each name in the ground truth according to one of the algorithms.

\subsubsection{Comparison to Phonetic Encoding Algorithms}
\label{sec:phonetic_encoding_algorithms_evaluation}
 
We evaluated the performance of five well-known phonetic algorithms: Soundex, Metaphone, Double Metaphone, NYSIIS, and the MRA (matching rating approach) for the task of suggesting synonyms.
The evaluation process was performed as follows:
First, for all of the names in WikiTree, all of the phonetic codes were calculated. 
Next, for each name in the ground truth, the phonetic code was returned.
For example, for the name John in the ground truth, the code J500 was retrieved (encoded by Soundex algorithm).
Then, we chose those names that shared the same phonetic code as John as candidates; we then sorted the candidates according to their edit distance from the given name (the lower the distance, the higher the similarity) and obtained the top $K$ as synonyms.

Unlike phonetic algorithms which produce a single sound code for a given name, Double Metaphone produces two phonetic codes (primary and secondary). 
Therefore, for this algorithm, we collected all of the names that shared the same phonetic code (as either the primary or secondary code) and ordered them according to their edit distance from the given name. 

The time complexity for these algorithms took into account the time to calculate all of the phonetic encoding codes for all of the names in the dataset, the time to obtain the candidates which share the same phonetic code, and the sorting time.

\subsubsection{Comparison to String Similarity Algorithms}
\label{sec:string_similarity_algorithms_evaluation}

We evaluated the performance of three well-known string similarity algorithms (edit distance, Damerau Levenshtein distance, and Jaro-Winkler distance).
For each algorithm, we measured the string similarity between each name in the ground truth and the candidate name existing in the WikiTree dataset.
Take, for example, the name John and the edit distance string similarity algorithm: 
First, we calculated the edit distance between each name in the WikiTree dataset and the name John.
%As candidates, we chose just the forenames whose distance from the given name is between one and three.
%We limited the edit distance so it was less than or equal to three, since we observed that a larger edit distance value resulted in names that were very different from the given name.
In the final step, we sorted the candidates based on their distance. 

In contrast to the first string similarity algorithms, the Jaro-Winkler distance ranges from zero to one.
So in this case, we sorted the candidates for a given name in descending order and chose just the top $K$ as synonyms.
Finally, to improve the performance, we sorted the $K$ synonyms according to their edit distance from the given name and retrieved them as synonyms. 

Here, the time complexity consisted of 1) the time to measure the similarity between each name in the ground truth and all other names in the dataset, and 2) the time required to sort the top 10 candidates.

\subsubsection{Comparison to Approximate String Joins with Abbreviations (ASJA)}
\label{sec:approximate_string_joins}
We also evaluated the performance of the algorithm suggested by Tao et al.~\cite{Tao2017ApproximateSJ}.
For this, we used their framework and provided it with two lists of unique forenames and surnames from WikiTree.
After the framework was executed, we filtered the names that did not exist in the ground truth dataset and selected the top 10 names suggested for evaluation.

The time complexity for this algorithm included the time required to suggest names for each name in the WikiTree dataset and the time it took to filter the names that did not exist in the ground truth.

\subsubsection{Comparison to Name2Vec}
\label{sec:name2vec_evaluation}

%In order to compare Name2Vec to \emph{GRAFT}, we evaluated the performance of the Name2Vec~\cite{foxcroft2019name2vec} algorithm, as well as \emph{GRAFT} on forenames and surnames.

\textbf{Evaluation of Name2Vec on Forenames.} 
We used the Name2Vec framework in order to train three Doc2Vec models based on names that exist in the WikiTree dataset. 
This means that we generated three Doc2Vec models, where the forenames and surnames that exist in the WikiTree dataset are viewed as documents in the two models, whereas the surnames in the Ancestry Surnames dataset served as document in the third model. 
During model training, we set the parameters so they were the same as those reported by Foxcroft et al.~\cite{foxcroft2019name2vec} (640 epochs, 30 dimensions, and a window size of two).
Next, using the trained model, we collected the 10 most similar candidate names for each name in the ground truth according to the cosine similarity distance.
Then, we used these candidates as synonyms for the given names. 

The time complexity for Name2Vec took into account the time for training the Doc2Vec model based on the names provided and the time required to suggest the most similar names for a given name in the ground truth.

% First, we used the WikiTree dataset as a data source and trained a Doc2Vec model based on these forenames.
% Foxcroft et al. reported that their best model training on the Ancestry Surnames dataset consisted of 250,000 surnames.
% Since the datasets are nearly equal in size (250,000 records), and there are generally no major differences between forenames and surnames, we set the parameters so they were the same as those reported by Foxcroft et al. (640 epochs, 30 dimensions, and a window size of two). 
% Next, using the trained model, we collected the 10 most similar candidate names for each name in the ground truth.
% Then, we used these candidates as synonyms for the given names. 

% \textbf{Evaluation of Name2Vec on Surnames.}
% In this experiment, we performed the same steps described in the previous paragraph:
% This time, we trained a Doc2Vec model with the parameters described above on the Ancestry Surnames dataset, which includes 250,000 surnames.
% Finally, the remaining candidates were evaluated using the Ancestry Records ground truth dataset. 

\subsection{Improving \emph{GRAFT}}
To determine the best parameters for improving \emph{GRAFT}, we evaluated the performance of \graft using different graph types, edit distance ranges for various thresholds and ordering functions. 
% To determine the best parameters for improving \emph{GRAFT}, we evaluated the performance of the proposed \graft algorithm based on the generation of different $G_N^{i,j}$ graphs, \bluecomment{the four ordering functions ($Similarity_i, i=1..4$)}, and the edit distance ranges for various thresholds.
In total, we executed 64 experiments, where the independent variables were four graph types (children--parents, grandchildren--grandparents, great-grandchildren--great-grandparents, and children--all ancestors), four ranges of edit distance for filtering the candidates (ranged from two to five) and four ordering functions. %($Similarity_i, i=1..4$).}
We used the WikiTree dataset to generate the graphs and the three ground truth datasets (BtN Forenames and Surnames datasets and the Ancestry Surnames dataset as the ground truth.

\subsection{Analysis of the Graph Size} 
In order to understand the effect of the size of constructed name-based graphs, we analyzed the constructed graphs' size according to a few edit distance ranges. 
In total, we measured the size of the graphs constructed, where the independent variables were the different graph types (children--parents, grandchildren--grandparents, great-grandchildren--great-grandparents, and children--all ancestors) and the edit distance for filtering the candidates ranged from one to five.
For forenames and surnames, we used the WikiTree dataset to generate the graphs.

\section{Results}
\label{sec:results}

\subsection{Performance Comparison}
\label{sec:performance_comparison}

To analyze the \graft algorithm with respect to other algorithms, we evaluated the performance of all algorithms using three ground truth datasets: the Behind the Name (BtN) Forenames and Surnames, and the Ancestry Surnames datasets (see Section~\ref{sec:data}).
% In this section, we present the results of the experiments described in Section~\ref{sec:evaluation}.
% The results of this evaluation are presented in Tables~\ref{tab:top_10_performance_wikitree_forenames}, \ref{tab:top_10_performance_btn_surnames}, and \ref{tab:top_10_performance_ancestry_surnames}.

First, we evaluated the performance of all algorithms on forenames (see Table~\ref{tab:top_10_performance_wikitree_forenames}).
We found that \graft when using the $G_N^{1,3}$ graph constructed based on children and their grandparents, with the $NetEDofDMphoneED$ ordering function obtained the highest performance scores with respect to average accuracy, F1, and precision at K (precision@k), where $k =1,2,3,5,10$.
It obtained average accuracy and F1 scores of 0.115 and 0.17, respectively.
%With respect to% 
Regrading the recall measure, \emph{HGRAFT}, the hybrid algorithm, obtained the highest average recall score of 0.27 (see Table~\ref{tab:top_10_performance_wikitree_forenames}).  

\begin{table*}[t]
  \centering
  \caption{Top 10 Performance on Behind the Name Forenames Dataset}
  \begin{tabular}{cccccccccccc}
Method & Accuracy & F1 & AP@1 & AP@2 & AP@3 & AP@5 & AP@10 & Recall & Time (sec) & Cover & Cover(\%)\\ [0.5ex] 
 \hline\hline
% $NetED$ & \bluecomment{0.185} & \bluecomment{0.245} & \bluecomment{0.349} & \bluecomment{0.275} & \bluecomment{0.241} & \bluecomment{0.211} & \bluecomment{0.185} & \bluecomment{0.18}\\
% $NetED$ & 0.11 & 0.162 & 0.275 & 0.209 & 0.181 & 0.149 & 0.11 &  0.167 & 749.1 & 1,260 & 20\% \\

% $Net^2ED$ & 0.107 & 0.158 & 0.261 & 0.203 & 0.172 & 0.14 & 0.107 & 0.162 & 800.38 & 1,260 & 20\% \\

% $EDofDMphone$ & 0.097 & 0.144 & 0.236 & 0.176 & 0.149 & 0.122 & 0.098 & 0.153 & 898.75 & 1,260 & 20\%\\

% $NetEDofDMphoneED$ & \textbf{0.115} & \textbf{0.17} & \textbf{0.294} & \textbf{0.225} & \textbf{0.191} & \textbf{0.154} & \textbf{0.115} & 0.174 & 1,023.52 & 1,260 & 20\%\\
GRAFT & \textbf{0.115} & \textbf{0.17} & \textbf{0.294} & \textbf{0.225} & \textbf{0.191} & \textbf{0.154} & \textbf{0.115} & 0.174 & 1,023.52 & 1,260 & 20\%\\
%parents-children
%$Similarity_4$ + DM & 0.076 & 0.123 & 0.09 & 0.079 & 0.072 & 0.067 & 0.076 & \textbf{0.223} \\
%grandparents-grandsons
\hline
% $NetEDofDMphoneED$ + DM & \textbf{\bluecomment{0.082}} & \textbf{\bluecomment{0.133}} & \textbf{\bluecomment{0.154}} & \textbf{\bluecomment{0.138}} & \textbf{\bluecomment{0.127}} & \textbf{\bluecomment{0.109}} & \textbf{\bluecomment{0.082}} & \textbf{\bluecomment{0.27}} & \bluecomment{1,474.71} &  \bluecomment{6,236} & \bluecomment{99.4\%} \\
% \hline
HGRAFT & 0.082 & 0.133 & 0.154 & 0.138 & 0.127 & 0.109 & 0.082 & \textbf{0.27} & 1,474.71 &  6,236 & 99.4\% \\
\hline

Soundex & 0.064 & 0.109 & 0.093 & 0.094 & 0.092 & 0.083 & 0.064 & 0.257 & 363.55 & 6,270 & 99.9\%\\

Metaphone & 0.066 & 0.117 & 0.104 & 0.102 & 0.099 & 0.089 & 0.069 & 0.261 & 354.37 & 6,232 & 99.3\% \\

DMetaphone & 0.07 & 0.118 & 0.104 & 0.103 & 0.1 & 0.09 & 0.07 & 0.268 & 411.42 & 6,236 & 99.3\%   \\

NYSIIS & 0.065 & 0.106 & 0.103 & 0.098 & 0.092 & 0.081 & 0.065 & 0.208 & 326.5 & 6,113 & 97.4\% \\ 

MRA & 0.055 & 0.087 & 0.078 & 0.078 &	0.077 &	0.071 &	0.055 & 0.151 & \textbf{314.19} & 6,009 & 95.7\% \\ 
\hline

Name2Vec & 0.023 & 0.042 & 0.051 & 0.042 & 0.037 & 0.032 & 0.023 & 0.096 & 18,055.6 & 6,273 &  99.9\% \\
\hline

%suggested to all the names
Jaro-Winkler & 0.043 & 0.076 &	0.074 &	0.069 &	0.064 & 0.055 & 0.043 & 0.197 & 8,027.485 & \textbf{6,274} & \textbf{100\%} \\ 

% % suggested only to names that starting with the same character
% Jaro-Winkler & \bluecomment{0.044} & \bluecomment{0.078} &	\bluecomment{0.076} &	\bluecomment{0.069} & \bluecomment{0.065} & \bluecomment{0.058} &	\bluecomment{0.044} & \bluecomment{0.203} & \bluecomment{759.12} & \bluecomment{6,274} & \bluecomment{100\%} \\ 

Edit Distance & 0.049 & 0.085 &	0.075 &	0.072 & 0.069 &	0.06 & 0.049 &	0.212 & 1,197.717 & \textbf{6,274} & \textbf{100\%} \\ 

DLD & 0.049 & 0.086 & 0.079 & 0.074 & 0.067 & 0.061 & 0.049 & 0.214 & 1,227.99 & \textbf{6,274} & \textbf{100\%} \\
\hline
ASJA & 0.015 & 0.024 & 0.025 & 0.019 & 0.018 & 0.017 & 0.015 & 0.037 & 1,495.53 & 6,004 & 95.7\% \\

  \end{tabular}
  \label{tab:top_10_performance_wikitree_forenames}
\end{table*}

Second, concerning  time complexity and coverage, we can see that the fastest algorithm was the matching rating approach (MRA) which took 314.19 seconds, whereas the string similarity algorithms (edit distance, DLD, and Jaro Winkler) suggested synonyms for all of the names that exist in the ground truth. 

Third, with respect to surnames evaluation (see Table~\ref{tab:top_10_performance_btn_surnames} and \ref{tab:top_10_performance_ancestry_surnames}), we found that \graft when using the $G_N^{1,3}$ graph constructed based on children and their parents, with the $NetEDofDMphoneED$ ordering function obtained the highest performance scores with respect to average accuracy, F1, and precision at K (precision@k), where $k =1,2,3,5,10$ on both the BtN and Ancestry Surnames datasets.
On the BtN Surnames dataset, \graft obtained average accuracy, F1, and precision@1 of 0.195, 0.25, and 0.367, respectively. 
On the Ancestry Surnames dataset, \graft obtained even higher results (average accuracy, F1, and precision@1 of 0.492, 0.532, and 0.631, respectively).
With respect to recall, it can be seen that on the BtN Surnames dataset, \graft obtained the highest average recall score of 0.19, whereas on the Ancestry Surnames dataset, Damerau-Levenshein Distance (DLD) obtained the highest average recall score of 0.894. 

\begin{table*}[t]
  \centering
  \caption{Top 10 Performance on Behind the Name Surnames Dataset}
  \begin{tabular}{ccccccccccccc}
Method & Accuracy & F1 & AP@1 & AP@2 & AP@3 & AP@5 & AP@10 & Recall & Time (sec) & Cover & Cover (\%)\\ [0.5ex] 
 \hline\hline
 
% GRAFT parents-children & \textbf{0.195} & \textbf{0.25} & \textbf{0.367} & \textbf{0.295} & \textbf{0.258} & \textbf{0.225} & \textbf{0.195} & \textbf{0.19} & 708.08 & 1,353 & 43.1\% \\
GRAFT p-c & \textbf{0.195} & \textbf{0.25} & \textbf{0.367} & \textbf{0.295} & \textbf{0.258} & \textbf{0.225} & \textbf{0.195} & \textbf{0.19} & 708.08 & 1,353 & 43.1\% \\

GRAFT gp-gc & 0.154 &  0.194 & 0.303 & 0.238 & 0.205 & 0.177 & 0.154 & 0.125 & 803.74 & 1,108 & 35.3\%\\
\hline
% GRAFT grandparents-children & 0.154 &  0.194 & 0.303 & 0.238 & 0.205 & 0.177 & 0.154 & 0.125 & 803.74 & 1,108 & 35.3\%\\
% \hline

% \bluecomment{GRAFT parents-children + NYSIIS} & \bluecomment{0.135} & \bluecomment{0.183} & \bluecomment{0.25} & \bluecomment{0.206} & \bluecomment{0.183} & \bluecomment{0.159} & \bluecomment{0.135} & \bluecomment{0.169} & \bluecomment{934.77} & \bluecomment{2,673} & \bluecomment{85.15\%}  \\
% \hline
HGRAFT & 0.135 & 0.183 & 0.25 & 0.206 & 0.183 & 0.159 & 0.135 & 0.169 & 934.77 & 2,673 & 85.15\%  \\
\hline

Soundex & 0.061 & 0.105 & 0.094 & 0.093 & 0.089 & 0.081 & 0.061 & 0.177 & 241.07 & 2,485 & 79.17\% \\

Metaphone & 0.059 & 0.099 & 0.102 & 0.093 & 0.087 & 0.077 & 0.059 & 0.158 & 222.87 & 2,462 & 78.43\% \\

DMetaphone & 0.059 & 0.1 & 0.102 & 0.093 & 0.087 & 0.077 & 0.059 & 0.16 & 251.01 & 2,463 & 78.46\% \\

NYSIIS & 0.07 & 0.113 & 0.111 & 0.105 & 0.096 &	0.086 & 0.07 & 0.152 & 212.73 & 2,399 & 76.43\% \\ 

MRA & 0.049 & 0.078 & 0.077 &	0.071 &	0.07 &	0.062 &	0.049 & 0.1 & \textbf{209.54} & 2,332 & 74.29\% \\ 
\hline

Name2Vec & 0.026 & 0.046 & 0.053 & 0.052 & 0.044 & 0.035 & 0.026 & 0.075 & 12,565.09 & \textbf{3,139} & \textbf{100\%} \\
\hline

% % For only the names that starting with the same word as original
% \bluecomment{Jaro-Winkler} & \bluecomment{0.056} & \bluecomment{0.096} &	\bluecomment{0.091} &	\bluecomment{0.088} &	\bluecomment{0.085} &	\bluecomment{0.073} & \bluecomment{0.056} & \bluecomment{0.169} & \bluecomment{308.57} & \bluecomment{3,139} & \bluecomment{X\%} \\ 

% Sוuggesting for all names
Jaro-Winkler & 0.053 & 0.092 &	0.089 &	0.083 &	0.079 &	0.068 & 0.053 & 0.159 & 3,224.11 & \textbf{3,139} & \textbf{100\%} \\ 

ED & 0.051 & 0.088 &	0.079 & 0.072 & 0.068 & 0.061 & 0.051 &	0.153 & 8,952.87 & \textbf{3,139} & \textbf{100\%}\\ 

DLD & 0.05 & 0.087 & 0.065 & 0.067 & 0.066 & 0.061 & 0.05 & 0.154 & 8,934.79 & \textbf{3,139} & \textbf{100\%} \\
\hline
ASJA & 0.024 & 0.038 & 0.031 & 0.028 & 0.027 & 0.025 & 0.024 & 0.043 & 466.28 & 3,037 & 96.75\%\\
  \end{tabular}
  \label{tab:top_10_performance_btn_surnames}
\end{table*}

\begin{table*}[t]
  \centering
  \caption{Top 10 Performance on Ancestry Surnames Dataset}
  \begin{tabular}{ccccccccccccc}
Method & Accuracy & F1 & AP@1 & AP@2 & AP@3 & AP@5 & AP@10 & Recall & Time (sec) & Cover & Cover (\%)\\ [0.5ex] 
 \hline\hline
 
GRAFT p-c & \textbf{0.492} & \textbf{0.532} & \textbf{0.631} & \textbf{0.537} & \textbf{0.511} & \textbf{0.498} & \textbf{0.492} & 0.66 & 820.34 & 3,084 & 13.07\% \\

GRAFT gp-gc & 0.38 &  0.4 & 0.5 & 0.416 & 0.397 & 0.386 & 0.38 & 0.504 & 858.5 & 1,918 & 8.13\%\\
\hline

% \bluecomment{GRAFT parents-children + NYSIIS} & \textbf{\bluecomment{0.309}} & \textbf{\bluecomment{0.383}} & \textbf{\bluecomment{0.563}} & \textbf{\bluecomment{0.447}} & \textbf{\bluecomment{0.391}} & \textbf{\bluecomment{0.343}} & \textbf{\bluecomment{0.309}} & \bluecomment{0.809} & \bluecomment{1,945.22} & \bluecomment{22,266} & \bluecomment{94.4\%}  \\
% \hline
HGRAFT & 0.309 & 0.383 & 0.563 & 0.447 & 0.391 & 0.343 & 0.309 & 0.809 & 1,945.22 & 22,266 & 94.4\%  \\
\hline

Soundex & 0.098 & 0.174 & 0.531 & 0.371 & 0.278 & 0.183 & 0.098 & 0.884 & 1,015.57 & 23,580 & 99.9\% \\

Metaphone & 0.164 & 0.241 & 0.513 & 0.376 & 0.3 & 0.225 & 0.164 & 0.831 & 955.52 & 23,038 & 97.67\% \\

DMetaphone & 0.16 & 0.237 & 0.515 & 0.376 & 0.298 & 0.222 & 0.16 & 0.839 & 1,175.88 & 23,084 & 97.86\% \\

NYSIIS & 0.266 & 0.346 & 0.562 & 0.43 & 0.364 &	0.306 &	0.266 &	0.835 & 894.7 & 22,192 & 94.08\% \\ 

MRA & 0.245 & 0.304 & 0.43 & 0.336 & 0.295 & 0.262 & 0.245 & 0.593 & 873 & 20,547 & 87.1\% \\ 
\hline

Name2Vec & 0.211 & 0.287 & 0.4 & 0.312 & 0.263 & 0.223 & 0.211 & 0.611 & 6,349.94 & 22,818 & 96.74\% \\
\hline

% For all Names
Jaro-Winkler & 0.086 & 0.155 &	0.347 &	0.263 &	0.21 &	0.148 & 0.086 & 0.815 & 11,475.07 & \textbf{23,588} & \textbf{100\%} \\ 

% % just strating with the same first word 
% \bluecomment{Jaro-Winkler} & \bluecomment{0.09} & \bluecomment{0.162} &	\bluecomment{0.38} &	\bluecomment{0.283} &	\bluecomment{0.224} &	\bluecomment{0.157} & \bluecomment{0.09} & \bluecomment{0.853} & \bluecomment{1,100.14} & \bluecomment{23,587} & \bluecomment{100\%} \\ 

ED & 0.093 & 0.167 &	0.341 & 0.272 & 0.223 & 0.161 & 0.093 &	0.874 & 1,197.37 & \textbf{23,587} & \textbf{100\%}\\ 

DLD & 0.095 & 0.171 & 0.348 & 0.277 & 0.228 & 0.164 & 0.095 & \textbf{0.894} & 1,742.32 & \textbf{23,587} & \textbf{100\%} \\
\hline
ASJA & 0.074 & 0.111 & 0.019 & 0.039 & 0.056 & 0.071 & 0.074 & 0.282 & \textbf{137.027} & 22,594 & 95.79\\
  \end{tabular}
  \label{tab:top_10_performance_ancestry_surnames}
\end{table*}

Fourth, regarding time complexity, on the Behind the Name Surnames dataset, like the forename case, the MRA was found to be the quickest (209.54 seconds).
On the Ancestry Surnames dataset, the quickest algorithm was ASJA at 137.027 seconds for suggesting synonyms for 22,594 names.

Finally, with respect to coverage, the algorithms obtaining the highest coverage were again the string similarity algorithms (edit distance, DLD, and Jaro-Winkler) which suggested synonyms for all of the names existing in the both surnames ground truth datasets.
In addition, Name2Vec on the Behind the Name Surnames dataset suggested synonyms for all of the names.

\subsection{Improving \graft}
\label{sec:improving_graft_results}

To explore different parameters that maximize the performance of \emph{GRAFT}, we evaluated the performance of the four ordering functions proposed ($Similarity_i, i=1..4$) with respect to different edit distance ranges and types of graphs. 

\begin{figure*}[h!]
\centering
\includegraphics[scale=0.6]{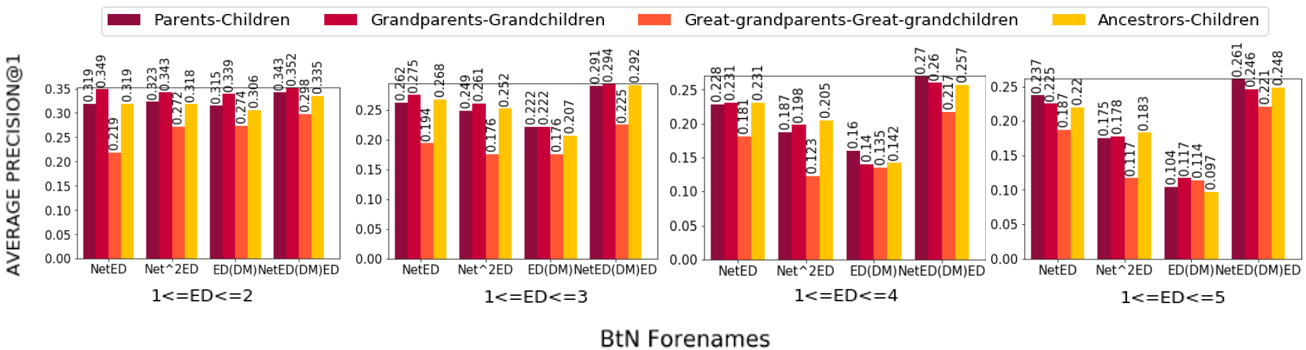}
\caption{\graft performance comparison based on different graphs of names generated with different ancestors, thresholds, and similarity functions evaluated on the Behind the Name (BtN) Forenames dataset.}
\label{fig:comparision_ordering_function_EDs_performance_evaluation_on_btn_forenames}
\end{figure*}

\begin{figure*}[h!]
\centering
\includegraphics[scale=0.6]{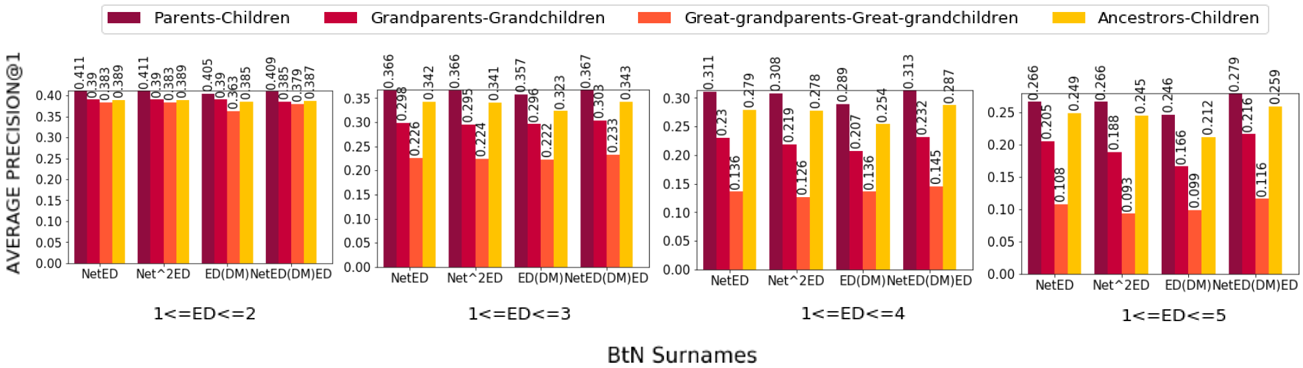}
\caption{\graft Performance comparison based on different graphs of names generated with different ancestors, thresholds, and similarity functions evaluated on the Behind the Name (BtN) Surnames dataset.}
\label{fig:comparision_ordering_function_EDs_performance_evaluation_on_btn_surnames}
\end{figure*}

\begin{figure*}[h!]
\centering
\includegraphics[scale=0.6]{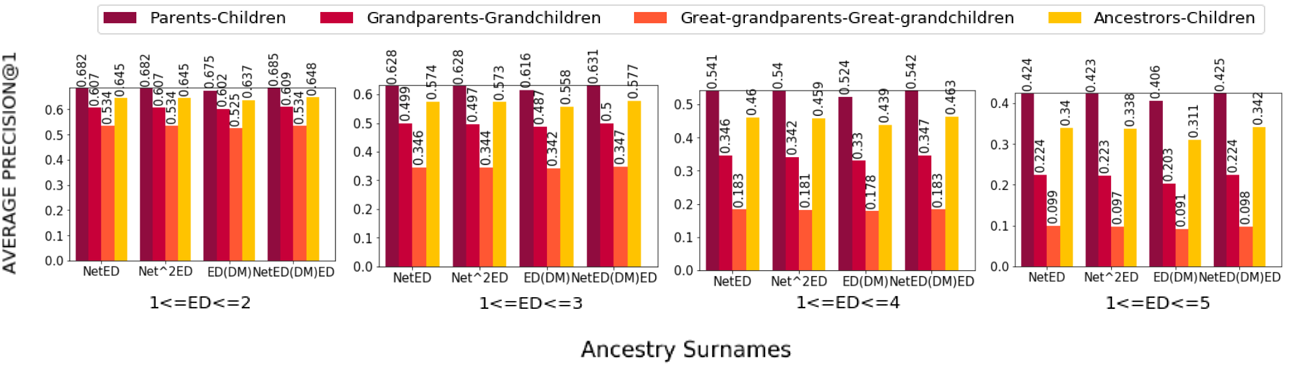}
\caption{\graft Performance comparison based on different graphs of names generated with different ancestors, thresholds, and similarity functions evaluated on the Ancestry Surnames dataset.}
\label{fig:comparision_ordering_function_EDs_performance_evaluation_on_ancestry_surnames}
\end{figure*}

We found that $NetEDofDMphoneED$ obtained the highest average precision@1 scores for all edit distance ranges on all of the datasets (see Figures~\ref{fig:comparision_ordering_function_EDs_performance_evaluation_on_btn_forenames}, \ref{fig:comparision_ordering_function_EDs_performance_evaluation_on_btn_surnames}, and \ref{fig:comparision_ordering_function_EDs_performance_evaluation_on_ancestry_surnames}).

The edit distance ranges from one to two obtained the highest average precision@1 scores in all of the datasets.

Finally, with respect to the type of graph, we found that graphs of forenames constructed based on grandchildren and their grandparents obtained the highest average precision@1 scores (0.352).
However, regarding surnames, graphs constructed based on surnames of children and their parents obtained the highest average precision@1 score (0.411 on the Behind the Name Forenames dataset and 0.685 on the Ancestry Surnames dataset).

\subsection{Graph Size Analysis}
\label{sec:graph_size_analysis_results}

\begin{figure*}[h!]
\centering
\includegraphics[scale=0.6]{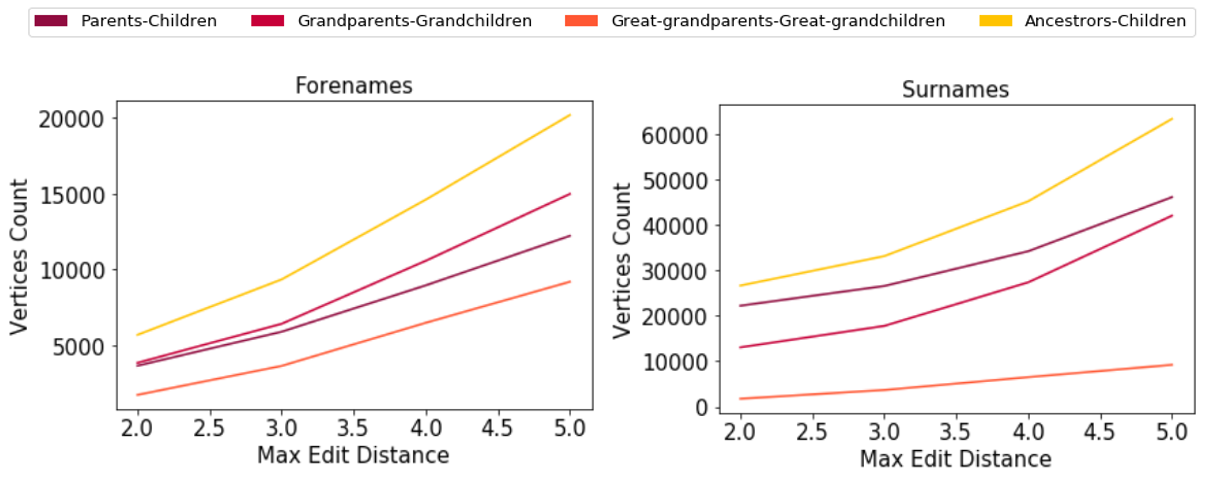}
\caption{Name-based graphs' size with respect to different edit distance ranges.}
\label{fig:comparision_nodes_count}
\end{figure*}

To understand the effect of the size of constructed name-based graphs, we examined the graph size based on the number of vertices comprising the name-based graph. 
First, we can see that the largest graphs constructed are based on all ancestors on both forenames and surnames.
Second, the size of graphs based on forenames with respect to edit distances of 1, 2, 3, 4, and 5 were comprised of 5,701, 9,354, 14,616, and 20,198 vertices, respectively.
Third, the size of graphs based on surnames with respect to edit distance of 1, 2, 3, 4, and 5 were comprised of 26,656, 33,143, 45,213, 63,370 vertices, respectively (see Figure~\ref{fig:comparision_nodes_count}).

\section{Discussion}
\label{sec:discussion}

Based on our analysis of the results presented in Section~\ref{sec:results}, we can conclude the following:
First, the proposed \graft algorithm was found useful for suggesting synonyms for both forenames and surnames (see Tables~\ref{tab:top_10_performance_wikitree_forenames}, \ref{tab:top_10_performance_btn_surnames}, and \ref{tab:top_10_performance_ancestry_surnames}).
\graft and \hgraft were found superior to all other algorithms evaluated, including encoding phonetic, string similarity algorithms, and Name2Vec which bases its suggestions on machine and deep learning.
We can conclude that the construction of a graph based on names derived from a genealogical dataset which utilizes generations of ancestral data, can be very effective for the synonym suggestion task.

Second, the \graft algorithm is generic in several respects.
For example, the name-based graph can be constructed using any genealogical dataset available. 
In this study, we used the WikiTree dataset, but any other dataset can be used.
Also, the name suggestion step can use any ordering function capable of ordering candidate names in a particular order.
We demonstrate this generic step using four ordering functions, but we certain that there are many more ordering functions that can improve performance.
Our demonstration of \graft on forenames and surnames also demonstrates the algorithm's generality.

%%%%%%%
Third, with respect to the proposed ordering functions, we can see that the $NetEDofDMphoneED$ function provided the best performance of the four ordering functions suggested; more specifically, it obtained the best performance for all of the types of graph constructed (grandchildren--grandparents, great-grandchildren--great-grandparents, and children--all ancestors), and for all of the edit distance ranges (see Figures~\ref{fig:comparision_ordering_function_EDs_performance_evaluation_on_btn_forenames}, \ref{fig:comparision_ordering_function_EDs_performance_evaluation_on_btn_surnames} and \ref{fig:comparision_ordering_function_EDs_performance_evaluation_on_ancestry_surnames}).
Therefore, we conclude that the use of ordering functions that consider several aspects, such as the graph structure, string similarity between the given and candidate names, as well as their phonetic codes, increases the likelihood of suggesting a larger number of relevant synonyms than other ordering functions that consider only one aspect. 
However, taking into account more aspects consumes more time, as well.
%very important for suggesting synonyms, however it is more time consuming.

Fourth, regarding the comparison of \graft and Name2Vec (see Section~\ref{sec:name2vec_evaluation}), we can see that \graft outperformed Name2Vec in every respect. 
% Regarding forenames, we can see that \graft which uses a graph based on names of grandparents and their grandchildren obtained the best results (see Table~\ref{tab:name2vec_comaprision}).
% Also, with respect to surnames, we can see that \graft with a graph based on names of parents and their children outperformed Name2Vec.
We conclude that the use of graphs based on names of children and their ancestors is much more effective for synonym suggestion than name embeddings generated based on the Doc2Vec methodology. 
Name2Vec refers to each name as a document and each letter constructing the name as a word in the document.
The problem with this approach is associated with the nature of embeddings which are greatly affected by the order of the words in each document. 
In cases in which there are many names, counting on the order of the letters will not result in satisfactory performance.

Fifth, the high performance of \graft and \hgraft demonstrates the effectiveness of these approaches for suggesting relevant synonyms for both forenames and surnames.
They were evaluated on three different ground truth datasets reflecting different domains associated with names. 
The Ancestry Surnames dataset consists of synonyms that share similar sound and characters (e.g., Philips and Phillips), whereas the Behind the Name (BtN) Forenames and Surnames datasets consist of personal names which share similar etymology (e.g., John and Johannes and Abrams and Abramsson).
%The success of \graft to suggest the highest number of correct synonyms for these datasets emphasizes its effectiveness.

Sixth, one of the \emph{GRAFT}'s challenges is related to its dependency on name-based graphs to suggest synonyms.
In cases in which a given name does not exist in the graph constructed, no candidate is suggested as a synonym.
Of course, when using large genealogical datasets, the likelihood of being unable to suggest synonyms is greatly reduced.
To address this, we proposed \emph{HGRAFT}, a hybrid algorithm which utilizes both \graft and a phonetic encoding algorithm for suggesting synonyms for all of the names (see Section~\ref{sec:hybrid_approach}).
It can be seen that \hgraft suggested synonyms for the majority of the names and contributed to superior performance in all respects (accuracy, F1, precision, and recall), an accomplishment not shared by any of the other algorithms evaluated. 
This finding demonstrates the \hgraft algorithm's effectiveness compared to other algorithms and emphasizes its ability to be used as both a standalone algorithm and in combination with other algorithms for suggesting names, in order to boost their capabilities.

Seventh, with respect to graph size and performance perspectives, we can see that while graphs constructed based on children and all of their ancestors (parents, grandparents, and great-grandparents) consisted of the greatest number of vertices, as opposed to other graphs, this was not matched by their performance, as they only achieved second place for both forenames and surnames (see Figure~\ref{fig:comparision_nodes_count}). %(second to graph based on grandparents and grandchildren for forenames suggestion and to graph based on parents and children for surnames suggestion).
In addition, it can be seen that much smaller graphs constructed based on forenames of grandchildren and their grandparents, as well as graphs constructed based on surnames of children and their parents, obtained the highest performance scores. 
Therefore, we can conclude the following: 
(a) It is recommended to use \graft based on a grandparents--grandchildren graph to suggest synonyms for forenames, and to use \graft based on a children--parents graph to suggest synonyms for surnames in order to achieve the best performance;
(b) these relatively small graphs (children--parents and grandparents--grandchildren graphs) compared to the very large children-all ancestors graphs, are much more distilled (i.e., are capable of capturing strong recurring pattern along family lines); 
(c) children are named after their grandparents much more often than their parents and great-grandparents, with respect to forenames (found statistically significant using t-tests with $p-value < 0.05$);
and (d) it can be seen that graphs constructed based on great grandparents--great-grandchildren, are largely small, compared to graphs constructed based on children and their parents, and graphs contructed based on grandparents and their grandchildren (see Figure~\ref{fig:comparision_nodes_count}).
In addition, the performance of graphs constructed based on great-grandparents--great-grandchildren was very low.
These results strengthen the conclusion of Rossi~\cite{rossi1965naming} who analyzed patterns of naming children in a sample of 347 urban middle-class mothers.
Rossi concluded that the kin for whom children were named are generally one or two generations (i.e., parents and grandparents and less frequently great-grandparents) removed from the child.

Eighth, concerning the threshold for filtering candidates, we can see that there is a tradeoff between edit distance ranges, graph size, and performance.
The larger the edit distance range, the poorer the performance (see Figures~\ref{fig:comparision_ordering_function_EDs_performance_evaluation_on_btn_forenames}, \ref{fig:comparision_ordering_function_EDs_performance_evaluation_on_btn_surnames}, and \ref{fig:comparision_ordering_function_EDs_performance_evaluation_on_ancestry_surnames}).
This is of course reasonable, since by increasing the edit distance’s range many irrelevant candidates are included as synonyms.
These irrelevant candidates are mainly the names of iconic biblical figures which have been popular throughout history, such as John, Ann and Mary, or names that are associated with one another in biblical stories but do not serve as synonyms, such as Abram and Sarah (husband and wife), Anna and Maria (mother and daughter), Mary and James (mother and son), and many more.
In the case of surnames, increasing the edit distance’s range results in many irrelevant candidates that share the same nationality (for example, Larson and Andersson, Martinez and Lopes) %or the same name by many people, such as Johnson, Smith, etc.  
or the same name being suggested for people with common, popular names, such as Johnson and Smith.

%In addition, it is important to understand that reducing the range too much will increase the performance, as well as reduce the number of names that the algorithm will be able to suggest as synonyms.

%%%%%%%%%%%%%%%%%%%%%%%%%%%%%%%%%%%%%%%%%%%%%%%%%

Ninth, with respect to limitations, the name-based graph is constructed based on similarity along family lines.
It is important to mention that \graft does not generalize to cultures in which people do not name their children after their relatives (by forename).
For example, many Chinese families usually give their new baby a name made up of two syllables from the Chinese alphabet consisting of thousands of characters, each with individual meanings~\cite{chiu30}.
As a result, it is rare to find two people with the same forename.\footnote{https://www.babycentre.co.uk/a568884/baby-naming-practices-from-around-the-world} %~\cite{baby_naming_practices}.
%Chinese forenames can theoretically include any of the Chinese language's 100,000 characters~\cite{chiu30}.
%In addition, in China and other Far Asian cultures, it is considered disrespectful to name a child after an older relative, and both bad practice and disadvantageous for the child's fortune to copy the names of famous historical figures.
Moreover, Some Asian surnames are shared by a vast number of people.
For example, the 14 most popular Vietnamese surnames account for the names of well over 90\% of Vietnamese population,\footnote{https://www.atlasobscura.com/articles/nguyen-name-common-vietnam} %~\cite{Vietnamese_last_names}, 
and the three Korean surnames of Kim, Lee, and Park accounted for almost half of the population of South Korea in 2014.\footnote{https://www.economist.com/the-economist-explains/2014/09/08/why-so-many-koreans-are-called-kim} %~\cite{korean_last_names}.  
This means that in such cases, the constructed graph probably would include a few relevant or many irrelevant candidates, either of which would result in decreased performance.
Therefore, we conclude that \graft works best at suggesting synonyms in cases there is a strong recurring pattern along family lines and in a phonetic language.

Finally, regarding performance, we can see that in general, all of the algorithms, including \emph{GRAFT}, obtained low accuracy scores with respect to both forenames and surnames (around 0.1 on the BtN Forenames and 0.2-0.3 on the Ancestry Surnames datasets, respectively). 
Based on these results, we conclude that suggesting synonyms for forenames and surnames is a task which is extremely difficult and for which there is a great room for improvement.
Therefore, we think that measures used in the information retrieval domain (e.g., precision@k) can shed some light on the performance and the ground truth.
With respect to to forenames, we can also see that the precision measures are very low.
We believe that the low results are related to the nature of the given dataset; when examining the datasets, we can see that many of the synonyms are associated with the etymology of the names which are very different from the original name.
For example, the synonym for Aabraham is Aapo, and the synonym of Alexandra is Sasha. 
These examples cannot be suggested by \graft or any of the other algorithms evaluated, since they are close historically but not semantically. 
On the other hand, \graft suggested synonyms that seemed at first sight as legitimate synonyms, but turned out not to be (e.g., Winfred and Alfred, Roberta and Berta, and Dana and Dina).  

Regarding the Ancestry Surnames dataset, we can see that the precision scores (obtaining an average precision@1 score of 0.68) are much higher than in the BtN Forenames dataset .
We believe that this is also related to the nature of the Ancestry Surnames ground truth dataset. 
This ground truth dataset consists of 25K pair names where the second name in each pair serves as a synonym for the first name. 
Ancestry defines a synonym for a given name to appear high number of occurrences in many family lines (e.g., Clark-Clarke, Parrish-Parish, and Seymour-Seymore), which can be more easily detected by \graft and phonetic encoding algorithms than detecting synonyms, using the BtN Forenames dataset which consists of many names that share a similar etymology (e.g., Alexander and Sasha) which are much harder to detect by most of the algorithms.

\section{Conclusion \& Future Work}
\label{sec:conclusion}

This paper introduces \emph{GRAFT}, a novel and generic algorithm which utilizes genealogical data to address some of the challenges associated with synonym suggestion. 
We provided a comprehensive description of the proposed algorithm’s steps, which start with the collection of a genealogical dataset.
After cleaning the data, we constructed digitized family trees for users of the dataset.
Based on these family trees, we constructed a name-based graph derived from children and their ancestors. 
Using this graph, as well as four ordering functions, we suggested synonyms.
To compare the results obtained, we evaluated the performance of 10 other algorithms, including phonetic and string similarity algorithms, as well as Name2Vec, a machine and deep learning approach for suggesting synonyms presented in 2019.
The evaluation was performed on forename and surname datasets.

We make the following observations and conclusions.
The proposed \graft algorithm outperformed all other algorithms evaluated, including encoding phonetic and string similarity algorithms, as well as Name2Vec.
We conclude that utilizing a genealogical dataset and network science was found very effective for synonym suggestion.
Also, \emph{GRAFT}'s success rates on three different datasets: the first two datasets of which focus on the etymology of forenames and surnames, whereas the third of which focuses on surnames that are seen in many family lines, emphasizes its usefulness.

The proposed algorithm is generic and a few aspects of its generality were demonstrated in our research, such as its ability to suggest synonyms for both forenames and surnames, as well as the ability to apply different ordering functions to improve the algorithm's performance.

Although \emph{GRAFT}'s dependency on a name-based graph to suggest synonyms, \emph{HGRAFT}, the hybrid algorithm which combines \graft and a phonetic encoding algorithm, was found superior to all other algorithms.
This finding emphasizes the hybrid algorithm's ability to suggest more accurate synonyms, as well as its ability to suggest synonyms for all names.    

With respect to fine-tuning, we recommend constructing a name-based graph derived from grandparents and their grandchildren to improve performance for the task of suggesting synonyms for forenames, whereas for surnames, we suggest constructing a parents--children graph.

Regarding limitations, it is important to emphasize that \graft works best in an environment where names share similarity along a high number of family lines. 
The algorithm's performance is not strong in cases, in which the dataset has less name diversity or names which are shared by many people. 
Future research directions could include examining other elements, such as gender, ethnicity, homeland, and other factors that affect name suggestion, nationality prediction, and nickname suggestion using digitized family trees and data extracted from the Web.
%Also, the sorting functions can be configured to assist in detecting aliases, or machine learning techniques could be used to improve the name suggestions.

\section{Availability}

\emph{GRAFT}'s synonym suggestions, as well as the three ground truth datasets collected are available upon request.
% This study is reproducible research. 
% Therefore, a code suggesting synonyms for a given name is available.\footnote{https://github.com/aviade5/GRAFT} 
% Other datasets for evaluation are available upon request. 

\section{Acknowledgments}
% Carol Teegarden for proofreading this article, and
The authors would like to thank the icons8 website (https://icons8.com) for their beautiful icons.

%\newpage\clearpage

\bibliographystyle{unsrt}
\bibliography{references.bib}
\newpage

\end{document}